\documentclass[12pt]{article}
\usepackage{amssymb,amsfonts,hyperref,graphicx,enumerate,color,bbold,wick,amsmath,physics}

\newcommand{\be}{\begin{equation}}
\newcommand{\ee}{\end{equation}}
\newcommand{\bea}{\begin{eqnarray}}
\newcommand{\eea}{\end{eqnarray}}
\newcommand{\beas}{\begin{eqnarray*}}
\newcommand{\eeas}{\end{eqnarray*}}

\newcommand{\identity}{\mathbb{1}}

\begin{document}
\begin{titlepage}

\medskip

\begin{center}

{\Large \bf Dissipation in the $1/D$ expansion\\for planar matrix models}

\vspace{10mm}

\renewcommand\thefootnote{\mbox{$\fnsymbol{footnote}$}}
Takanori Anegawa${}^{1}$\footnote{takanegawa@gmail.com},
Norihiro Iizuka${}^{2,3}$\footnote{iizuka@phys.nthu.edu.tw}, 
Daniel Kabat${}^{4,5}$\footnote{daniel.kabat@lehman.cuny.edu}

\vspace{6mm}

${}^1${\small \sl Yonago College, National Institute of Technology}\\
{\small \sl Yonago, Tottori 683-8502, Japan}

\vspace{3mm}

${}^2${\small \sl Department of Physics, National Tsing Hua University} \\
{\small \sl Hsinchu 30013, Taiwan}

\vspace{3mm}

${}^3${\small \sl Yukawa Institute for Theoretical Physics} \\ 
{\small \sl Kyoto University, Kyoto 606-8502, Japan}

\vspace{3mm}

${}^4${\small \sl Department of Physics and Astronomy} \\
{\small \sl Lehman College, City University of New York} \\
{\small \sl 250 Bedford Park Blvd.\ W, Bronx NY 10468, USA}

\vspace{3mm}

${}^5${\small \sl Graduate School and University Center, City University of New York} \\
{\small \sl  365 Fifth Avenue, New York NY 10016, USA}

\end{center}

\vspace{12mm}

\noindent
{We consider the thermal behavior of a large number of matrix degrees of freedom in the planar limit.  We work in $0+1$ dimensions,
with $D$ matrices, and use $1/D$ as an expansion parameter.  This can be thought of as a non-commutative large-$D$ vector model, with two independent
quartic couplings for the two different orderings of the matrices.  We compute a thermal two-point correlator to ${\cal O}(1/D)$ and find that the degeneracy present
at large $D$ is lifted, with energy levels split by an amount $\sim 1/\sqrt{D}$.  This implies a timescale for thermal dissipation $\sim \sqrt{D}$.  At high temperatures dissipation
is predominantly due to one of the two quartic couplings.}

\end{titlepage}
\setcounter{footnote}{0}
\renewcommand\thefootnote{\mbox{\arabic{footnote}}}

\hrule
\tableofcontents
\bigskip
\hrule

\addtolength{\parskip}{8pt}
\section{The model}
Consider the quantum mechanics of a collection of $N \times N$ Hermitian matrices $X^i(\tau)$, $i = 1,\ldots,D$.
We describe them using a Euclidean action
\be
\label{SE}
S = \int d\tau \, {1 \over 2} {\rm Tr} \, \big(\partial_\tau X^i \partial_\tau X^i\big) + {1 \over 2} m_0^2 \, {\rm Tr} \, \big(X^i X^i\big)
+ {1 \over 2} g_A^2 \, {\rm Tr} \, \big(X^i X^i X^j X^j\big) - {1 \over 2} g_C^2 \, {\rm Tr} \, \big(X^i X^j X^i X^j\big)
\ee
Here $m_0$ is a bare mass parameter, and we've introduced two quartic couplings $g_A$, $g_C$.

We will study the model (\ref{SE}) for its own sake.  However as motivation note that if we set $m_0 = 0$ and $g_A = g_C = g_{YM}$
then the action reduces to
\be
\label{SE2}
S = \int d\tau \, {1 \over 2} {\rm Tr} \, \big(\partial_\tau X^i \partial_\tau X^i\big) - {1 \over 4} g_{YM}^2 \, {\rm Tr} \, \big([X^i,X^j]^2\big)
\ee
Although the model we consider has no gauge symmetry, the same potential term appears in the dimensional reduction of $U(N)$
Yang-Mills theory from $D+1$ to $0+1$ dimensions.  The commutator-squared potential is also familiar in the BFSS matrix model
\cite{Banks:1996vh}.  In the model (\ref{SE}) we treat $g_A$ and $g_C$ as independent couplings since, as we will see, they lead to rather
different dynamics.

We're interested in the leading large-$N$ limit, in which only planar diagrams contribute.  However we're also interested
in the behavior for large $D$.  So instead of holding the two 't Hooft couplings fixed, we instead consider the limit
\bea
\nonumber
& & \lambda_A = g_A^2 N \rightarrow 0 \quad {\rm with} \quad \tilde{\lambda}_A = \lambda_A D \quad {\rm fixed} \\[5pt]
\label{scaling}
& & \lambda_C = g_C^2 N \rightarrow 0 \quad {\rm with} \quad \tilde{\lambda}_C = \lambda_C D \quad {\rm fixed}
\eea

Our goal is to study dissipation in this model at large $D$.  That is, we're interested in dissipation in a many-matrix model.
This is a tractable problem because the model has an $SO(D)$ symmetry that acts
on the $i,j$ indices, and from that point of view it's similar to a large-$D$ vector model and we can use $1/D$ as an expansion parameter.  However from the $U(N)$
point of view we're restricting to planar diagrams, which means it's {\it not} a standard vector model.  Instead (\ref{SE}) defines a
sort of {\it non-commutative} vector model, which lets us distinguish between the two couplings $g_A$ and $g_C$.  Another perspective
on the model is to think of $X^i_{AB}$ as a three-index object, which means we are dealing with a tensor model \cite{Klebanov:2018fzb}
in a particular scaling limit.  A different scaling limit was considered in \cite{Ferrari:2017ryl, Azeyanagi:2017drg}. In the literature the $1/D$ expansion has been developed to study
correlation functions \cite{Hotta:1998en} and the thermal partition function \cite{Mandal:2009vz}, and it has been applied to a commuting vector model in \cite{Kolganov:2022mpe}.
Related techniques were used to study Lyapunov exponents in scalar field theory in \cite{Stanford:2015owe}. 

\section{Hubbard-Stratonovich approach}
\label{Hubbard-Stratonovich}
\subsection{Hubbard-Stratonovich transformation}
To proceed it's convenient to perform a Hubbard-Stratonovich transformation and introduce an auxiliary Hermitian field $\Sigma$.
\begin{align}
\label{SE3}
S &= \int d\tau \, {1 \over 2} {\rm Tr} \, \big(\partial_\tau X^i \partial_\tau X^i\big) + {1 \over 2} m_0^2 \, {\rm Tr} \, \big(X^i X^i\big)
+ {1 \over 2} {\rm Tr} \, \big(\Sigma^2\big) - i g_A {\rm Tr} \, \big(\Sigma X^i X^i\big) \nonumber \\
& \qquad  \quad - {1 \over 2} g_C^2 \, {\rm Tr} \, \big(X^i X^j X^i X^j\big)
\end{align}
The Gaussian path integral over $\Sigma$ is well-defined, and the saddle point fixes $\Sigma = i g_A X^i X^i$.  Integrating out $\Sigma$ using its algebraic equation of motion recovers (\ref{SE}).\footnote{A more careful argument is given in appendix \ref{appendix:auxiliary}.}  This is a standard step for large-$D$ vector models and as we will see it's
the most convenient way to treat the coupling $g_A$.  We set
\be
\label{define}
\Sigma = \Sigma_0 \, \identity_{N \times N} + \sigma \qquad\quad m^2 = m_0^2 - 2 i g_A \Sigma_0
\ee
so that
\bea
\nonumber
S & = & \int d\tau \, {1 \over 2} {\rm Tr} \, \big(\partial_\tau X^i \partial_\tau X^i\big) + {1 \over 2} m^2 \, {\rm Tr} \, \big(X^i X^i\big)
+ {1 \over 2} {\rm Tr} \, \big(\sigma^2\big) + \Sigma_0 {\rm Tr} \, \big(\sigma\big) - i g_A {\rm Tr} \, \big(\sigma X^i X^i\big) \\
\label{SE4}
& & \hspace{1cm} - {1 \over 2} g_C^2 \, {\rm Tr} \, \big(X^i X^j X^i X^j\big)
\eea
We choose the parameter $\Sigma_0$ so that the vev of $\sigma$ vanishes.  This fixes\footnote{This condition is
a consequence of one of the Schwinger - Dyson equations of the model (\ref{SE4}), namely
$\langle {\delta S \over \delta \sigma} \rangle = 0$ or equivalently $\langle \sigma \rangle = i g_A \langle X^i X^i \rangle - \Sigma_0 \, \identity_{N \times N}$.}
\be
\label{tadpole}
\Sigma_0 \, \identity_{N \times N} = i g_A \, \langle X^i X^i \rangle
\ee
or equivalently
\be
\label{tadpole2}
\Sigma_0 = i g_A \, {1 \over N} \langle {\rm Tr} \big(X^i X^i\big) \rangle
\ee
With $\Sigma_0$ fixed in this way, (\ref{define}) gives an equation that fixes the mass $m$ of the fields $X^i$.  Note that
$m$ has been defined so that the vev of $\sigma$ vanishes.  At large $D$ it is also the location of the pole in the $X^i$ propagator,
but as we will see $1/D$ corrections to the propagator shift the location of the pole.  So in general $m$ is simply a parameter that
characterizes the theory.

Before proceeding it's worth doing some dimensional analysis.  For the action to be dimensionless we have
\bea
\nonumber
&& X^i \sim ({\rm mass})^{-1/2} \\
&& \sigma,\,\Sigma_0 \sim ({\rm mass})^{1/2} \\
\nonumber
&& g_A^2,\,\lambda_A,\,\tilde{\lambda}_A,\,g_C^2,\,\lambda_C,\,\tilde{\lambda}_C \sim ({\rm mass})^3
\eea

\subsection{Zero temperature results\label{sect:2D}}
\begin{figure}
\centerline{\includegraphics{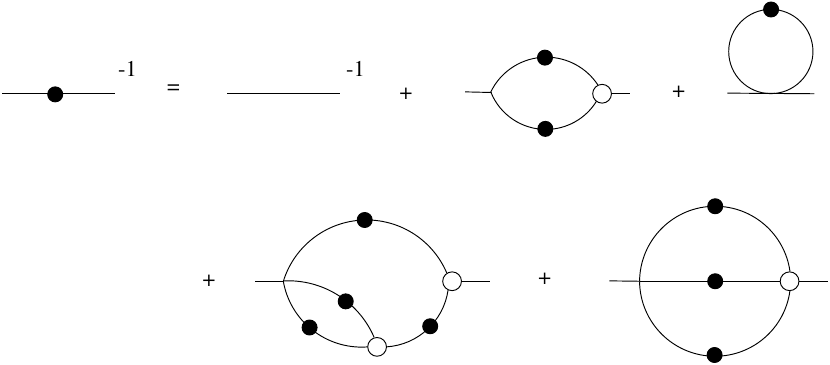}}
\caption{Schematic form of the Schwinger-Dyson equations for a theory with 3-point and
4-point couplings but no tadpoles.  Solid blobs are dressed
propagators; empty circles are 1PI vertices.  In the loop diagrams all numerical factors have been suppressed and
all external lines are understood to be amputated.  See for example \cite{Bjorken:1965sts,Roberts:1994dr,vonSmekal:1997ern}.
Figure taken from \cite{Kabat:1999hp}.\label{fig:SDeqns}}
\end{figure}

We begin by studying the 2-point functions in the model (\ref{SE4}) at zero temperature.  We do this by self-consistently
solving the Schwinger-Dyson equations of the model to ${\cal O}(1/D)$.  The condition (\ref{tadpole}) eliminates tadpoles,
and as a result, the Schwinger-Dyson equations we need to solve are schematically shown in Fig.~\ref{fig:SDeqns}.

The diagrams in Fig.~\ref{fig:SDeqns} are schematic in the sense that numerical factors have been suppressed.  To proceed
we introduce bare propagators (with $U(N)$ and $SO(D)$ indices suppressed)
\bea
&& \raisebox{-2mm}{\includegraphics[height=1cm]{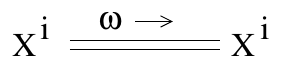}} \hspace{1cm} {\cal B}_X(\omega) = {1 \over \omega^2 + m^2} \\[5pt]
&& \,\, \raisebox{-2mm}{\includegraphics[height=1cm]{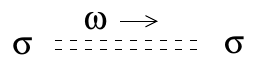}} \hspace{1.2cm} {\cal B}_\sigma(\omega) = 1
\eea
and dressed propagators
\bea
&& \raisebox{-2mm}{\includegraphics[height=1cm]{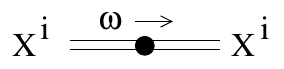}} \hspace{1cm} {\cal D}_X(\omega) \\[5pt]
&& \,\raisebox{-2mm}{\includegraphics[height=1cm]{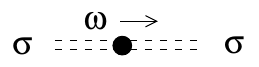}} \hspace{1.2cm} {\cal D}_\sigma(\omega)
\eea
These are related through 1PI self-energies by
\bea
&& {\cal D}_X^{-1} = {\cal B}_X^{-1} - {\cal E}_X \\[5pt]
&& {\cal D}_\sigma^{-1} = {\cal B}_\sigma^{-1} - {\cal E}_\sigma
\eea

\begin{figure}
\centerline{\includegraphics[width=16.5cm]{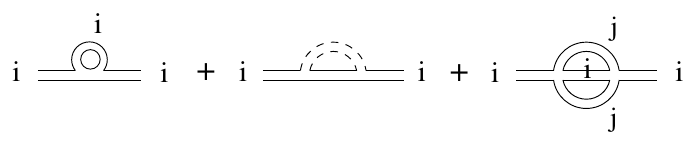}}
\caption{\label{fig:SDX}
Diagrams that contribute to ${\cal E}_X$ to ${\cal O}(1/D)$.  Internal lines are understood to be dressed propagators, external lines are amputated.
There's no sum on $i$, but there is a sum on $j$ in the last diagram.  In double-line notation the first two diagrams also appear in the flipped forms \raisebox{-7mm}{\protect\includegraphics{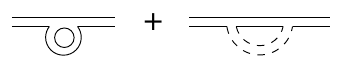}}.}
\end{figure}

\begin{figure}
\centerline{\includegraphics[width=14cm]{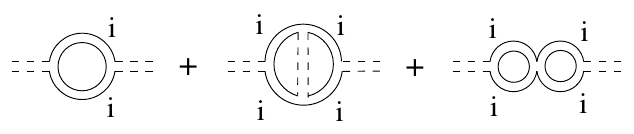}}
\caption{Diagrams that contribute to ${\cal E}_\sigma$ to ${\cal O}(1/D)$.  Internal lines are understood to be dressed propagators, external lines
are amputated.  There's a sum over the vector index $i$.\label{fig:SDsigma}}
\end{figure}

Using Fig.~\ref{fig:SDeqns} as a guide, the diagrams that contribute to ${\cal E}_X$ to ${\cal O}(1/D)$ are shown in
Fig.~\ref{fig:SDX}.  This leads to
\bea
\nonumber
D_X^{-1}(\omega) & = & \omega^2 + m^2 - \Big[4 g_c^2 N \int {dk \over 2\pi} D_X(k) \\
\nonumber
& & - 2 g_A^2 N \int {dk \over 2\pi} D_X(k + \omega) D_\sigma(k) \\
\label{DXinv}
& & + 4 g_C^4 N^2 D \int {dk_1 \over 2\pi} {dk_2 \over 2\pi} D_X(k_1) D_X(k_2) D_X(k_1 + k_2 + \omega) \Big]
\eea
(for an explanation of numerical factors see appendix \ref{appendix:combinatorics}).
Likewise the diagrams that contribute to ${\cal E}_\sigma$ to ${\cal O}(1/D)$ are shown in Fig.~\ref{fig:SDsigma} and
lead to
\bea
\nonumber
D_\sigma^{-1}(\omega) & = & 1 - \Big[-g_A^2 N D \int {dk \over 2\pi} D_X(k) D_X(k + \omega) \\
\nonumber
& & + 2 g_A^4 N^2 D \int {dk_1 \over 2\pi} {dk_2 \over 2\pi} D_X(k_1) D_X(k_2) D_\sigma(k_1 - k_2) D_X(k_1 + \omega)
D_X(k_2 + \omega) \\
& & - 2 g_A^2 g_C^2 N^2 D \int {dk_1 \over 2\pi} {dk_2 \over 2\pi} D_X(k_1) D_X(k_2) D_X(k_1 + \omega) D_X(k_2 + \omega)\Big]
\eea

Since we have $1/D$ as an expansion parameter it's quite easy to solve these equations.  At leading order for large $D$,
the only contribution to the self-energies comes from the first term (the bubble term) in ${\cal E}_\sigma$.  This means that at leading order
for large $D$ we have
\bea
\nonumber
D_\sigma^{-1}(\omega) & = & 1 + \tilde{\lambda}_A \int {dk \over 2\pi} \, {1 \over k^2 + m^2} \, {1 \over (k + \omega)^2 + m^2} \\
& = & 1 + {\tilde{\lambda}_A \over m (\omega^2 + 4 m^2)}
\eea
Thus
\be
\label{Dsig}
D_\sigma = {\omega^2 + 4 m^2 \over \omega^2 + m_\sigma^2}
\ee
where we have defined
\be
\label{msig}
m_\sigma^2 = 4 m^2 + \tilde{\lambda}_A / m
\ee

We use this result to evaluate the loop integrals in the $X^i$ self-energy to ${\cal O}(1/D)$.  The integrals in (\ref{DXinv}) lead to
\begin{align}
\label{zeroTpropagator}
D_X^{-1}(\omega) &= \omega^2 + m^2 - {4 \tilde{\lambda}_C \over 2 m D} + {\tilde{\lambda}_A \over m D} \, {\omega^2 + 5 m^2 + m m_\sigma + 4 m^3 / m_\sigma \over \omega^2 + (m + m_\sigma)^2} - {3 \tilde{\lambda}_C^2 \over m^2 D} \, {1 \over \omega^2 + 9 m^2}
\end{align}
We can evaluate the self-energy on-shell, by setting the Euclidean momentum to $\omega^2 = - m^2$.  We see that
the self-energy is real and to ${\cal O}(1/D)$ the only effect is a small shift in the physical (pole) mass of the fields $X^i$.

We can also determine the relation between the parameter $m$ and the bare parameters of the model (\ref{SE}).  The tadpole condition (\ref{tadpole}) fixes
\be
\Sigma_0 = i g_A N D \int {dk \over 2\pi} \, D_X(k)
\ee
so that
\be
\label{fullmdef}
m^2 = m_0^2 + 2 \tilde{\lambda}_A \int {dk \over 2\pi} \, D_X(k)
\ee
At leading order for large $D$ we have $D_X(k) = 1/(k^2 + m^2)$ so that
\be
\label{massallorder}
m^2 = m_0^2 + \tilde{\lambda}_A / m + \order{1/D}
\ee
It's useful to think of the bare mass as a function of $m$ by writing
\be
\label{mass2}
m_0^2 = m^2 - \tilde{\lambda}_A / m + \order{1/D}
\ee
Neglecting $ \order{{1/D}}$ corrections, the entire range $- \infty < m_0^2 < \infty$ corresponds to $m^2 > 0$, so we always have a positive dressed mass even
if the bare fields are tachyonic.  From (\ref{massallorder}) we see that $m^2 > m_0^2$ and from (\ref{msig}) we see that $m_\sigma^2 > 4 m^2$.

It is the leading large-$D$ behavior of the dressed mass that will be most relevant for us, especially in section \ref{direct}.  We denote this leading large-$D$ behavior by
$m_1$, with
\be
\label{mass1}
m_1^2 = m_0^2 + \tilde{\lambda}_A / m_1 
\ee
Thus $m_1$ agrees with $m$ up to $ \order{{1/D}}$ corrections. 

\subsection{Finite temperature results\label{sect:higher}}
We now study the model at finite temperature, with the goal of understanding dissipation in the $1/D$ expansion.
We do this using a Euclidean formalism, by discretizing the loop integrals to Matsubara sums.
\bea
\label{Matsubara}
\omega \rightarrow \omega_n = {2 \pi n \over \beta} \\
\int {d\omega \over 2\pi} \rightarrow {1 \over \beta} \sum_n
\label{discretesum}
\eea
We'll repeat the steps in the previous section: first determine the $\sigma$ propagator at leading order for large $D$, then
determine the $X^i$ propagator to ${\cal O}(1/D)$.

For $\sigma$ this gives the leading-order propagator
\be
D_\sigma^{-1}(\omega_n) = 1 + \tilde{\lambda}_A {1 \over \beta} \sum_n {1 \over k_n^2 + m^2} {1 \over (k_n + \omega_n)^2 + m^2}
\ee
Following the standard Saclay technique \cite{Pisarski:1987wc,Parwani:1991gq,Laine:2016hma} we switch to writing the propagators in position space,
\be
{1 \over \omega^2 + m^2} = \int_0^\beta d\tau \, e^{i \omega \tau} {1 \over 2m} \left[(1 + N_m) e^{-m\tau} + N_m e^{m\tau} \right]
\ee
where $N_m = {1 \over e^{\beta m} - 1}$ is a Bose distribution.  Then we do the Matsubara sum using
\be
{1 \over \beta} \sum_n e^{i 2 \pi n (\tau - \tau') / \beta} = \sum_w \delta(\tau - \tau' - \beta w)
\ee
The delta function kills one of the integrals over Euclidean time, and the remaining time integral leads to
\be
\label{thermalsigma}
D_\sigma^{-1}(\omega_n) = 1 + {\tilde{\lambda}_A \over (2m)^2} N_m^2 \left[{4 m \over \omega_n^2 + 4 m^2} \left(e^{2\beta m} - 1\right) + 2 \beta e^{\beta m} \delta_{n,0}\right]
\ee
The last term, proportional to $\delta_{n,0}$, comes from the cross terms between $e^{-m\tau}$ and $e^{+m\tau}$.
We take the inverse to get the propagator itself.  This gives a rather complicated expression which for convenience
we write in the form
\be
\label{thermalsigma2}
D_\sigma(\omega_n) = 1 - {m_\sigma^2 - 4 m^2 \over \omega_n^2 + m_\sigma^2} - A \delta_{n,0}
\ee
Here we have defined a thermally-corrected $\sigma$ mass
\be
m_\sigma^2 = 4 m^2 + {\tilde{\lambda}_A \over m} {e^{2\beta m} - 1 \over \big(e^{\beta m} - 1\big)^2}
\ee
We have also introduced a parameter $A$ to obtain the correct propagator for the zero mode.  It's fixed by requiring
that (\ref{thermalsigma}) and (\ref{thermalsigma2}) are consistent when $n = 0$.
\be
\label{Adef}
\Big[\hbox{\rm $D_\sigma(\omega_{n = 0})$ from (\ref{thermalsigma})}\Big] = {4 m^2 \over m_\sigma^2} - A
\ee
The parameter $A$ is a way of accounting for the $\delta_{n,0}$ term in (\ref{thermalsigma}).  Note that the $\delta_{n,0}$
term makes a positive contribution to the right hand side of (\ref{thermalsigma}), so it decreases the value of $D_\sigma(\omega_{n = 0})$, which means that $A$ is positive.

Next we evaluate the $X^i$ propagator to ${\cal O}(1/D)$.  In terms of Matsubara sums we have
\bea
\nonumber
D_X^{-1}(\omega) & = & \omega^2 + m^2 - 4 {\tilde{\lambda}_C \over D} {1 \over \beta} \sum_n {1 \over k_n^2 + m^2} \\
\nonumber
& & + 2 {\tilde{\lambda}_A \over D} {1 \over \beta} \sum_n {1 \over (k_n + \omega)^2 + m^2} D_\sigma(k_n) \\
& & - 4 {\tilde{\lambda}_C^2 \over D} {1 \over \beta^2} \sum_{n_1,n_2} {1 \over k_{n_1}^2 + m^2} {1 \over k_{n_2}^2 + m^2}
{1 \over (k_{n_1} + k_{n_2} + \omega)^2 + m^2}
\eea
The advantage of writing the $\sigma$ propagator in the form (\ref{thermalsigma2}) is that the Matsubara sums are all straightforward, following the steps used to obtain (\ref{thermalsigma}).  To ${\cal O}(1/D)$ we find
\bea
\nonumber
& & D_X^{-1}(\omega) = \omega^2 + m^2 + {2 \tilde{\lambda}_A - 4 \tilde{\lambda}_C \over D} S_1
- {2 \tilde{\lambda}_A \over D} (m_\sigma^2 - 4 m^2) S_2 \\
\label{thermalX}
& & \hspace{2cm} {\color{magenta} - {2 \tilde{\lambda}_A \over D} {A \over \beta} {1 \over \omega^2 + m^2}}
- {4 \tilde{\lambda}_C^2 \over D} S_3
\eea
where the sums (valid for Matsubara frequencies, $\omega \in {2 \pi \over \beta} {\mathbb Z}$) are\footnote{The expression for $S_2$ is valid provided $m \not= m_\sigma$, which is the case in our model. 
If one sets $m = m_\sigma$ in (\ref{S2}) there is an additional contribution to the sum proportional to $\delta_{n,0}$ that can be seen in (\ref{thermalsigma}).}
\bea
& & S_1 = {1 \over \beta} \sum_n {1 \over k_n^2 + m^2} = {1 \over 2 m \tanh (\beta m / 2)} \\[5pt]
\label{S2}
& & S_2 = {1 \over \beta} \sum_n {1 \over (k_n + \omega)^2 + m^2} {1 \over k_n^2 + m_\sigma^2} \\
\nonumber
& & \qquad = {N_m N_{m_\sigma}
\over 4 m m_\sigma} \Big[\left(e^{\beta(m + m_\sigma)} - 1\right) \left({1 \over i \omega + m_\sigma + m} - {1 \over i \omega - (m_\sigma + m)}\right) \\
\nonumber
& & \hspace{2.7cm} + \left(e^{\beta m_\sigma} - e^{\beta m}\right) \left({1 \over i \omega + m_\sigma - m} - {1 \over i \omega - (m_\sigma - m)}\right) \Big] \\[5pt]
& & S_3 = {1 \over \beta^2} \sum_{n_1,n_2} {1 \over k_{n_1}^2 + m^2} {1 \over k_{n_2}^2 + m^2} {1 \over (k_{n_1} +
k_{n_2} + \omega)^2 + m^2} \\
\nonumber
& & \qquad = {N_m^3 \over (2m)^3} \Big[\left(e^{3\beta m} - 1\right) \left({1 \over i \omega + 3m} - {1 \over i \omega - 3 m}\right) \\
\nonumber
& & \hspace{2.3cm} {\color{magenta} + 3 e^{\beta m} \left(e^{\beta m} - 1\right) \left({1 \over i \omega + m} - {1 \over i \omega - m}\right)} \Big]
\eea

Our goal is to study dissipation.  To this end, since we are working in Euclidean space, we examine the behavior of the propagator in the vicinity of $\omega^2 + m^2 = 0$.  Most of the loop corrections in (\ref{thermalX}) are small and make
an ${\cal O}(1/D)$ shift in the location of the pole, but two terms (highlighted in magenta) are dangerous since they diverge at $\omega = \pm i m$.
Retaining just the dangerous terms we approximate the inverse propagator as
\be
\label{thermalX2}
D_X^{-1}(\omega) = \omega^2 + m^2 - {B \over \omega^2 + m^2}
\ee
where
\be
\label{Bdef}
B = {2 \tilde{\lambda}_A \over D} {A \over \beta} + {3 \tilde{\lambda}_C^2 \over Dm^2} N_m^2 e^{\beta m}
\ee
In terms of the parameters of the model, the explicit expression for $B$ that follows from (\ref{Adef}) and (\ref{Bdef}) is  
\begin{align}
B = {3 \tilde{\lambda}_C^2 \over 4 D m^2 \sinh^2 {\beta m \over 2}}
- {2 \tilde{\lambda}_A \over D \beta} \left[
{1 \over 1 + {\tilde{\lambda}_A \over {4} m^3} \left({1 \over \tanh {\beta m \over 2}} + {\beta m \over 2 \sinh^2 {\beta m \over 2}}\right)}
- {1 \over 1 + {\tilde{\lambda}_A \over {4} m^3 \tanh{\beta m \over 2}}}\right] 
\end{align}

Taking the inverse of (\ref{thermalX2}), the propagator is
\be
D_X(\omega) = {1 \over 2} \left({1 \over \omega^2 + m^2 + \sqrt{B}} + {1 \over \omega^2 + m^2 - \sqrt{B}}\right)
\ee
Recall that $A$ is positive, which means that $B$ is also positive.  Also note that $A$ is ${\cal O}(1)$ while $B$ is
${\cal O}(1/D)$.  So we see that at finite temperature the single pole at $\omega^2 + m^2 = 0$ is split into a pair of
nearby poles at $\omega^2 + m^2 \pm \sqrt{B} = 0$.

To see the physical consequences we turn to the retarded Green's function $D_R(\omega)$, which can be obtained
from a Euclidean correlator by analytically continuing \cite{Laine:2016hma,2001CMaPh.216...59C}\footnote{Our conventions for Wick rotating are
$\omega_{\rm \scriptstyle Minkowski} = + i \omega_{\rm \scriptstyle Euclidean}$ and $t_{\rm \scriptstyle Minkowski} = - i t_{\rm \scriptstyle Euclidean}$.}
\be
\label{DR}
D_R(\omega) = D_X(\omega_n \rightarrow -i (\omega + i\epsilon))
\ee
This leads to
\be
D_R(\omega) = {1 \over 2} \left({1 \over - (\omega + i\epsilon)^2 + m_+^2} + {1 \over - (\omega + i\epsilon)^2 + m_-^2}\right)
\ee
where
\be
m_\pm^2 = m^2 \pm \sqrt{B}
\ee
At finite temperature effectively there are two nearby energy levels.  The consequences
are clearest if we transform back to position space, where
\bea
\nonumber
D_R(t) & = & \int {d\omega \over 2\pi} \, e^{-i \omega t} D_R(\omega) \\
\nonumber
& = & \theta(t) \, {1 \over 2} \left({1 \over m_+} \sin(m_+ t) + {1 \over m_-} \sin(m_- t)\right) \\
\label{DR2}
& \approx & \theta(t) \, {1 \over m} \sin(mt) \cos \Big({\sqrt{B} \, t \over 2 m}\Big)
\eea
The model behaves as though it were a discrete quantum mechanical system.  Although there is no true dissipation, the
two nearby energy levels lead to destructive interference on a timescale given by
\be
\tau = {\pi m \over \sqrt{B}}
\ee
We define an effective width for the excitation by
\be
\label{Gammadef}
\Gamma = {1 \over \tau} = {\sqrt{B} \over \pi m}
\ee
This is exponentially suppressed at low temperatures, where
\be
B \sim e^{-\beta m} \qquad \Gamma \sim e^{-\beta m / 2}
\ee
In contrast the width grows linearly at high temperatures, where
\be
\label{hiT}
\Gamma = {\sqrt{3} \, \tilde{\lambda}_C \over \pi m^3 \sqrt{D} \, \beta} + {\cal O}(\beta)
\ee

Finally we examine thermal corrections to the relation between the parameter $m$ and the bare parameters of the model (\ref{SE}).  At finite temperature the tadpole condition (\ref{tadpole}) fixes
\be
\Sigma_0 = i g_A N D \, {1 \over \beta} \sum_n D_X(k_n)
\ee
thus, from \eqref{define},
\be
m^2 = m_0^2 + 2 \tilde{\lambda}_A \, {1 \over \beta} \sum_n D_X(k_n)
\ee
At leading order for large $D$ we have $D_X(k) = 1/(k^2 + m^2)$ so that
\be
m^2 = m_0^2 + {\tilde{\lambda}_A \over m \tanh(\beta m / 2)} + {\cal O}(1/D)
\ee
We can think of the bare mass as a function of $m$ by writing
\be
m_0^2 = m^2 - {\tilde{\lambda}_A \over m \tanh(\beta m / 2)} + {\cal O}(1/D)
\ee
The entire range $- \infty < m_0^2 < \infty$ corresponds to $m^2 > 0$, so we always have a positive dressed mass even
if the bare fields are tachyonic.  This suggests there is no phase transition in the model.  

\section{Direct approach\label{direct}}

The same results can be obtained by directly analyzing the action \eqref{SE}.

\subsection{Zero temperature results}
\subsubsection{Leading order in $1/D$\label{sect:leading1/D}}
The two-point function for $X_{ij}$ field is 
\begin{align}
\ev{ X_{ij}(\tau) X_{kl}(0) }  = G(\tau) \delta_{il} \delta_{jk}
\end{align} 
and its Fourier transformation is
\begin{align}
G(\tau) = \int \frac{d\omega}{2\pi} \, G(\omega) \, e^{-i \omega \tau} \,, \quad G(\omega) = \int d\tau \, G(\tau) \, e^{i\omega \tau} \,.
\end{align}

The bare propagator is $G_0(\omega)$, and the Schwinger-Dyson equation for the dressed propagator $G(\omega)$ in the leading order in $1/D$ expansion becomes
\begin{align}
\label{largeDleadingG}
  G_0 = \frac{1}{\omega^2 + m_0^2} \,, \quad  G = \frac{1}{\omega^2 + m_1^2}  \,, \quad
G(\omega) = G_0(\omega) - c_L \tilde{\lambda}_A G_0(\omega) G(\omega) \int \frac{d\omega'}{2\pi} G(\omega')   \,
\end{align}
\begin{figure}[t]
\centering
\includegraphics[keepaspectratio,scale=0.4]{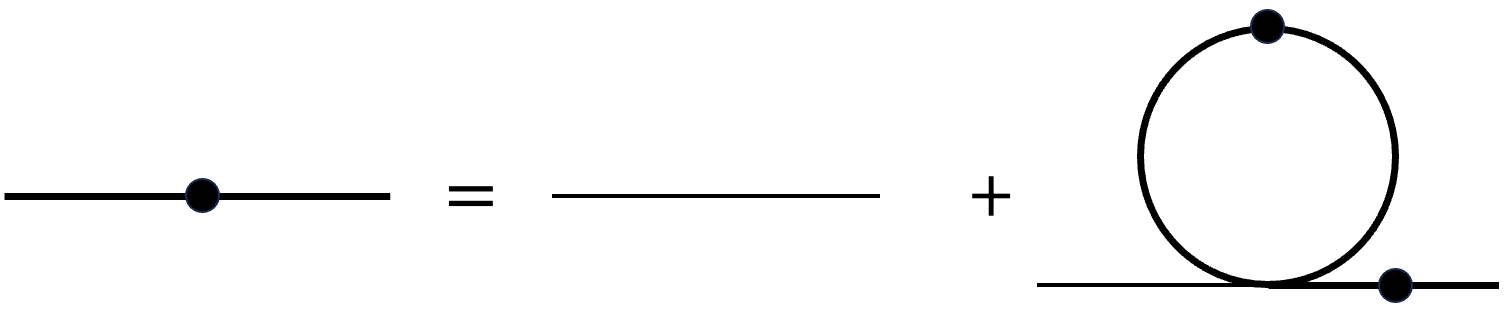}
\caption{Schwinger-Dyson equation in the leading order of $1/D$ expansion. In this leading order, only the snail diagrams contribute.}
\label{snailfig}
\end{figure}
This is obtained from the snail Feynman diagrams shown in Fig.~\ref{snailfig}. Here, $c_L$ is a constant determined by counting Wick contractions, and it turns out
$c_L=2$ (see Appendix \ref{app:leadingc}).
The 1-loop integral can be performed 
\begin{align}
\int d\omega' G(\omega') = \int d\omega' \frac{1}{\omega'^2 + m_1^2} 
=  \frac{\pi}{m_1}
\end{align}
and thus, 
\begin{align}
G_0(\omega)^{-1} = G(\omega)^{-1} - \frac{ \tilde{\lambda}_A}{m_1} \, \quad \Rightarrow \quad
\label{lmsftz}
 m_1^2 = m_0^2 +  \frac{\tilde{\lambda}_A}{m_1}   \,
\end{align}
This matches with \eqref{mass1}.

\subsubsection{A bubble chain diagram for $1/D$ corrections\label{bubblechainsection}}
\begin{figure}[htbp]
    \centering
    \includegraphics[keepaspectratio, scale=0.455]{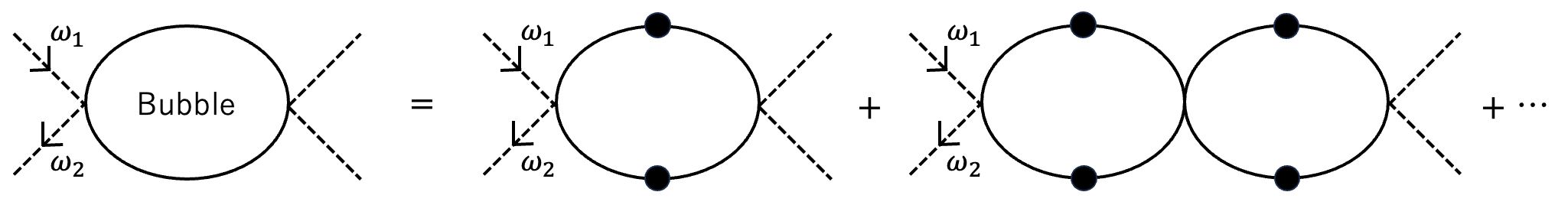}
    \caption{Bubble chain diagram with a momentum flow $\omega_1 - \omega_2$. The black circle propagator represents $G(\omega)$ which we already obtained in \eqref{lmsftz}.}
    \label{fig:Bubble}
\end{figure}

To evaluate $1/D$ corrections, we will have to take into account diagrams with a ``bubble chain.''
The bubble chain Feynman diagrams are defined in Fig.~\ref{fig:Bubble}. We denote them by $B(\omega)$, where $B(\omega)$ is the amplitude of the bubble chain diagrams such that there is a momentum flow $\omega = \omega_1 - \omega_2 \equiv \omega_{1-2}$.

From Fig.~\ref{fig:Bubble}, in Euclidean signature $B(\omega)$ satisfies
\begin{align}
B(\omega_{1-2}) = \int \frac{d\omega}{2\pi} G(\omega)G(\omega_{1-2}-\omega) -c_B \lambda_A D \int \frac{d\omega}{2\pi} G(\omega)G(\omega_{1-2}-\omega)B(\omega_{1-2}) \,.
\end{align}
Here $c_B$ is a coefficient. 
This shows that the bubble chain diagrams have a geometric sum structure as 
\begin{align}
\label{geometricformBomega}
B(\omega_{1-2}) = \frac{\int \frac{d\omega}{2\pi} G(\omega)G(\omega_{1-2}-\omega)}{1+ c_B  \tilde{\lambda}_A  \int \frac{d\omega}{2\pi} G(\omega)G(\omega_{1-2}-\omega)} \,
\end{align}
In Appendix \ref{app:bubblec}, we count Wick contractions and show that $c_B=1$. 
One can compute the integral using  
\begin{align}
G(\tau) = \frac{1}{2m_1} (\theta(\tau) e^{-m_1\tau} + \theta(-\tau) e^{m_1\tau})
\end{align}
as 
\begin{align}
& \int \frac{d\omega}{2\pi} G(\omega)G(\omega_{1-2}-\omega) 
= \int d \tau G(\tau)^2 e^{i\omega_{1-2} \tau} \nonumber \\
&\quad =  \frac{1}{m_1}  \int d \tau \frac{1}{4m_1}  (\theta(\tau) e^{-2m_1 \tau} + \theta(-\tau) e^{2m_1 \tau})e^{i\omega_{1-2} \tau} 
= \frac{1}{m_1} \frac{1}{\omega_{1-2}^2 + 4m_1^2}
\end{align}
From this, $B(\omega)$ can be determined as
\begin{align}
\label{Bomegazero}
B(\omega) 
= \frac{1}{m_1} \frac{1}{\omega^2 +  m_{\sigma}^2} \,, \quad \mbox{where} \quad
m_{\sigma}^2 = 4 m_1^2 + \frac{ \tilde{\lambda}_A}{m_1}  \, 
\end{align}
This expression for $m_\sigma$ agrees with \eqref{msig} up to $\order{1/D}$ corrections coming from the difference between $m$ and $m_1$.
Likewise the bubble chain propagator $B(\omega)$ is related to $D_\sigma(\omega)$ obtained in \eqref{Dsig} as 
  \begin{align}
1 - \tilde{\lambda}_A B(\omega) = 1 - \frac{\tilde{\lambda}_A}{m_1}\frac{1}{\omega^2 + m_\sigma^2}  = 1 - \frac{m_\sigma^2 - 4m_1^2 }{\omega^2 + m_\sigma^2} = D_\sigma(\omega)
\end{align}
again up to $\order{1/D}$ corrections coming from the difference between $m$ and $m_1$. 
For $\tilde{\lambda}_A > 0$ note that $m_{\sigma} >  2 m_1$.  For future reference the Fourier transform of $B(\omega)$ is
\begin{align}
B(\tau) 
& = \frac{1}{2 m_1 m_{\sigma}}  \left( \theta(\tau) e^{-m_{\sigma} \tau} + \theta(-\tau) e^{m_{\sigma} \tau} \right)  \,
\end{align}

\subsubsection{$\tilde{\lambda}_A/D$ corrections for dissipation}
Now we can evaluate the $1/D$ corrections to a two-point function which are responsible for dissipation. We will only be interested in diagrams that introduce
$\omega$ dependence.  There are additional self-energy diagrams that shift the mass; we ignore these diagrams for now and return to them in section \ref{sect:mass}. 

\begin{figure}[tbp]
    \centering
    \includegraphics[keepaspectratio, scale=0.7]{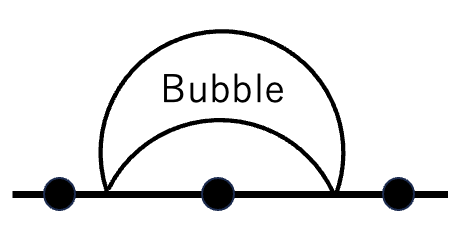}
  \caption{The self-energy diagram proportional to $\tilde{\lambda}_A/D$ which contributes to dissipation. The black circle propagators are dressed propagators $G(\omega)$ which we already obtained in \eqref{lmsftz} and do not include any $1/D$ corrections.  ``Bubble'' means the bubble chain diagram from Fig.~\ref{fig:Bubble}. This diagram
  makes an $\omega$-dependent contribution.}
  \label{fig:1/Dcorrection}
\end{figure}

There are two types of $1/D$ corrections, one is proportional to $\tilde{\lambda}_A/D$ and the other is  proportional to $\tilde{\lambda}_C/D$. 
We first consider only the contribution of $\tilde{\lambda}_A$. In other words, we set $\tilde{\lambda}_C = 0$ for a moment. Then the Feynman diagram proportional to
$\tilde{\lambda}_A/D$ that contributes to dissipation, {\it i.e.}, which produces a pole, is shown in Fig.~\ref{fig:1/Dcorrection}.  

Again we denote $G(\omega)$ as the leading dressed correlator in the large $D$ limit, {\it i.e.,} the propagator without $1/D$ correction given by \eqref{largeDleadingG}. 
Its mass $m_1$ is determined by \eqref{lmsftz} in the zero-temperature limit. 
We denote $\widetilde{G}(\omega)$ as the dressed correlator including $1/D$ corrections. The Schwinger-Dyson equation taking into account  $\tilde{\lambda}_A/D$ corrections becomes
\begin{align}
\widetilde{G}(\omega)^{-1} &= G(\omega)^{-1}  - \frac{c_A \tilde{\lambda}_A^2}{D}  \int \frac{d\omega_1}{2\pi} G(\omega_1) B(\omega - \omega_1)  
+ \delta m_A \,+ \, \order{\frac{1}{D^2}} 
\end{align}
Here we are focusing on the self-energy correction shown in Fig.~\ref{fig:1/Dcorrection}, since as we will see it has a pole.  In addition $\delta m_A$ represents the
$\omega$-independent $\order{\tilde{\lambda}_A/D}$ contributions coming from tadpole diagrams that are responsible for the $\order{\tilde{\lambda}_A/D}$ mass shift, {\it i.e.,} they are parts of the mass difference $m-m_1$. 
$c_A $ is a combinatoric constant that has the value $c_A= 2$, as shown in Appendix \ref{app:typea}.

Since 
\begin{align}
G(\tau) B(\tau) = \frac{m_1 + m_\sigma}{2 m_1^2 m_\sigma}  \frac{1}{2 (m_1 + m_\sigma)}\left( \theta(\tau) e^{-(m_1 + m_{\sigma}) \tau} + \theta(-\tau) e^{(m_1 + m_{\sigma} ) \tau} \right)
\end{align}
we obtain 
\begin{align}
\int \frac{d\omega_1}{2\pi} G(\omega_1) B(\omega - \omega_1) 
= \int d\tau G(\tau) B(\tau) e^{i \omega \tau} =   \frac{m_1 + m_\sigma}{2 m_1^2 m_\sigma}  \frac{ 1}{\omega^2 + (m_1 + m_{\sigma})^2 } 
\end{align}
Thus, 
\begin{align}
- \frac{2 \tilde{\lambda}^2_A}{D}  \int \frac{d\omega_1}{2\pi} G(\omega_1) B(\omega - \omega_1) 
&= 
- \frac{ \tilde{\lambda}_A^2}{D}    \frac{m_1 + m_\sigma}{ m_1^2 m_\sigma}
\frac{ 1}{\omega^2 + (m_1+m_{\sigma})^2} \nonumber \\
& \equiv \frac{1}{D} \Pi_{A} (\omega)
 \label{A1contribution}
\end{align}
The fact that it has a pole at $m_1 + m_{\sigma}$ is important.

Thus, the Schwinger-Dyson equation becomes
  \be
  \label{zeroTSDeqma}
\widetilde{G}(\omega)^{-1} = \omega^2 + m_1^2 + \frac{1}{D} \Pi_A(\omega) + \delta m_A + \order{\frac{1}{D^2}}
  \ee
where 
\begin{align}
\label{Pizero}
\frac{1}{D}\Pi_A(\omega) 
&= 
-  \frac{\tilde{\lambda}_A}{D}  \frac{\tilde{\lambda}_A}{m_1}    \frac{m_1 + m_\sigma}{ m_1 m_\sigma}
\frac{ 1}{\omega^2 + (m_1+m_{\sigma})^2}  
\\
& =   \frac{\tilde{\lambda}_A}{D}  \frac{1}{m_1 m_\sigma} 
  \frac{ \left(4 m_1^2 - m_\sigma^2  \right) \left( m_1 + m_\sigma \right)}{ \omega^2 + (m_1+m_{\sigma})^2 }
\end{align}
In the second equality, we use \eqref{Bomegazero} that relates $\tilde{\lambda}_A$ by $m_\sigma$ and $m_1$. 
This equation can be compared with $\tilde{\lambda}_A$ term in \eqref{zeroTpropagator}, where 
\begin{align}
\label{DanslambdaA}
 {\tilde{\lambda}_A \over m D} \, {\omega^2 + 5 m^2 + m m_\sigma + 4 m^3 / m_\sigma \over \omega^2 + (m + m_\sigma)^2}
& = \frac{\tilde{\lambda}_A}{D} \frac{1}{m m_\sigma} \left(  m_\sigma + \frac{ \left(4 m^2 - m_\sigma^2  \right) \left( m + m_\sigma \right)  }{\omega^2 + (m + m_\sigma)^2}   \right) 
 \end{align}
Thus the pole and its residue match completely.  We also see certain $\order{1/D}$ contributions to the mass difference $m-m_1 =\order{1/D}$. 

\subsubsection{$\tilde{\lambda}_C/D$ contributions for dissipation}
For a complete computation of the $1/D$ correction to a two-point function, we must incorporate the effect of $\tilde{\lambda}_C = g_C^2 N D$ as well. Then the Schwinger-Dyson equation becomes
\be
\label{zeroTSDeqmamc}
\widetilde{G}(\omega)^{-1} = \omega^2 + m_1^2 + \frac{1}{D}\left( \Pi_A(\omega) + \Pi_C(\omega) \right) + \delta m_A  + \delta m_C + \order{\frac{1}{D^2}}
\ee
\begin{figure}[tbp]
    \centering
    \includegraphics[keepaspectratio, scale=0.6]{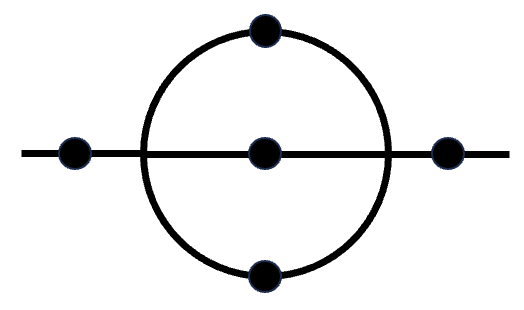}
    \caption{Melon diagram which contribute in $\tilde{\lambda}_C/D$ corrections.}
    \label{Melon}
\end{figure}
Here the melon diagram of Fig.~\ref{Melon} contributes to the self-energy as 
  \begin{align}
\frac{1}{D}\Pi_{C}(\omega) &= - \frac{c_C \tilde{\lambda}_C^2}{D} \int \frac{d \omega_1}{2 \pi } \int \frac{d \omega_2}{2 \pi } G(\omega_1)G(\omega_2)G(\omega-\omega_1 - \omega_2)\\
&=  - \frac{c_C \tilde{\lambda}_C^2}{D}  \int d\tau_1 \left( G(\tau_1) \right)^3 e^{i \omega \tau_1}\\
&= - \frac{3 c_C \tilde{\lambda}_C^2}{4 D m_1^2}   \frac{1}{\omega^2 + 9m_1^2}   
  \end{align}
  where $c_C$ is a combinatoric constant and $c_C = 4$ which we show in Appendix \ref{app:typec}.  
Comparing to the term proportional to $\tilde{\lambda}_C / D$ in \eqref{zeroTpropagator}, we see that the pole and its residue match completely.  We also see certain
$\order{\tilde{\lambda}_C/D}$ contributions to the mass difference $m - m_1$.

Note that there is also a melon diagram proportional to $\tilde{\lambda}_A = g_A^2 N D$. This melon diagram is obtained by replacing the bubble chain diagram in Fig.~\ref{fig:1/Dcorrection} by the first term on the right hand side of Fig.~\ref{fig:Bubble}. However note that the results from the two melon diagrams are very different. The melon diagram in Fig.~\ref{Melon} proportional to $\tilde{\lambda}_C$ gives a pole at $\omega^2 = - (3 m_1)^2$. On the other hand,  the melon diagram in Fig.~\ref{fig:1/Dcorrection} proportional to $\tilde{\lambda}_A$ gives a pole at $\omega^2 = - (m_1 + m_\sigma)^2$.

\subsubsection{$1/D$ contributions to the mass shift\label{sect:mass}}

Recall that we have introduced two slightly different definitions of dressed mass.
In section \ref{Hubbard-Stratonovich}, $m$ is defined by the no-tadpole condition which leads to \eqref{fullmdef}.
In section \ref{sect:leading1/D} we defined a leading large-$D$ dressed mass $m_1$ by solving a Schwinger-Dyson equation which leads to \eqref{mass1}.
The two definitions \eqref{fullmdef} and \eqref{mass1} differ by $\order{1/D}$ corrections.

In the direct approach of section \ref{direct}, there are many diagrams that are responsible for $\order{1/D}$ corrections to the mass.  We should take into account these $\order{1/D}$ mass shift corrections for comparison with the results in section \ref{Hubbard-Stratonovich}. 

For example, the snail diagrams in Fig.~\ref{snailfig} make a subleading $\order{1/D}$ correction to the mass. It comes from the following Wick contraction.
\be
+ \frac{ g^2_A}{2} \times 4 \times \sum_{j, k=1}^D \wick{121}{ <x X^i ( >x X^j <y X^k <z X^k >z X^j ) >y X^i}  
\ee
which gives part of the $\order{1/D}$ mass shift, $\delta m_A$, in \eqref{zeroTSDeqma} (or in \eqref{zeroTSDeqmamc}) as 
\begin{align}
\delta m_A =  {2 \tilde{\lambda}_A \over D} \int {d\omega_1 \over 2\pi} G(\omega_1)  + \cdots =  \frac{\tilde{\lambda}_A}{D m_1} + \cdots
\end{align}
This mass shift can be compared with the $\omega$-independent term in \eqref{zeroTpropagator} or equivalently in \eqref{DanslambdaA}.

Similarly, 
there is an additional mass shift contribution of $\order{\tilde{\lambda}_C/D}$, which is the snail diagram in Fig.~\ref{snailfig} but it is obtained by the $\lambda_C$ term where all indices are the same. More explicitly, 
\be
- \frac{ g^2_C}{2} \times 4 \times 2 \times \sum_{j=1}^D \wick{121}{ <x X^i (>x X^j <y X^j <z X^j >z X^j ) >y X^i }  
\ee
4 is because of 4 $X^j$'s and 2 is because of two choices for $\wick{1}{ <x X^j >x X^j}$. 
The self-energy from such a diagram gives 
  \begin{align}
\delta m_C &= - \frac{4 \tilde{\lambda}_C}{D}   \int \frac{d \omega}{2 \pi } G(\omega) + \cdots = - \frac{4 \tilde{\lambda}_C}{2 m_1 D}  + \cdots \,
  \end{align}
Again this mass shift can be compared with the $\omega$-independent term in \eqref{zeroTpropagator}.

\begin{figure}[tbp]
    \centering
    \begin{minipage}[b]{0.32\linewidth}
 \includegraphics[keepaspectratio, scale=0.6]{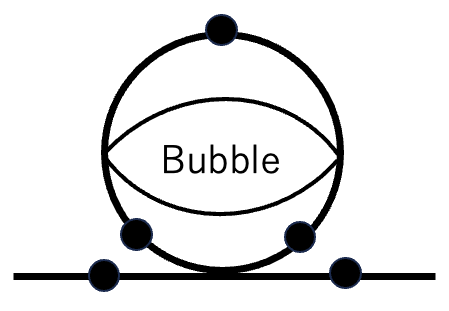}
  \end{minipage}
  \begin{minipage}[b]{0.32\linewidth}
    \includegraphics[keepaspectratio, scale=0.54]{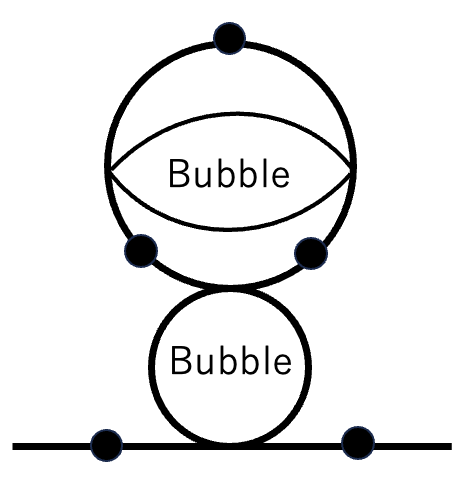}
  \end{minipage}
  \caption{Examples of diagrams which contribute to the mass shift at order $\tilde{\lambda}_A/D$. These tadpole diagrams are $\omega$-independent.  They contribute
  to the mass shift at $\order{1/D}$.}        \label{fig:massshift}
\end{figure}

In addition to these $\order{1/D}$ snail diagrams, many other diagrams are responsible for the mass shift. Examples of such diagrams are shown in Fig.~\ref{fig:massshift}. Since our main interests are in dissipation in many matrix models, we will not evaluate these mass shift contributions any further.

\subsection{Finite temperature results}
\subsubsection{Leading order in $1/D$}
At finite temperature, momentum is discretized as in \eqref{Matsubara} by Matsubara frequency and the integral becomes summation as in \eqref{discretesum}. The Schwinger-Dyson equation in the leading order in $1/D$ becomes  
\begin{align}
G(\omega_n) = G_0(\omega_n) - c_L \frac{\tilde{\lambda}_A}{\beta} G_0(\omega_n) G(\omega_n)\sum_k G(\omega_k) 
\end{align}
with $c_L = 2$. 
This becomes
\begin{align}
m_1^2 = m_0^2 + \frac{ \tilde{\lambda}_A}{m_1}\coth \frac{\beta m_1}{2}
\end{align}
In zero temperature limit $m_1 \beta \gg 1$, this becomes 
\begin{align}
m_1^2 = m_0^2 + \frac{ \tilde{\lambda}_A}{m_1}
\end{align}
which matches with \eqref{mass1}. 
 
On the other hand, in high temperature limit $m_1 \beta \ll 1$, this becomes
\begin{align}
m_1^2 \sim \frac{\tilde{\lambda}_A}{m_1} \frac{2}{\beta m_1}  = \frac{2 \tilde{\lambda}_A}{\beta m_1^2}
\end{align}

\subsubsection{A bubble chain diagram for $1/D$ corrections}

Let us perform a similar analysis for nonzero temperature. The analysis becomes complicated, but physically, the effective mass changes in finite temperature. 
First, let us see the bubble chain diagram for nonzero temperature. We can rewrite the Schwinger-Dyson equation as one with finite temperature.
\begin{align}
B((\omega_{1-2})_n) &= \frac{1}{\beta}\sum_k G(\omega_k)G((\omega_{1-2})_n- \omega_k) \nonumber \\
&\quad - \frac{c_B \lambda_A D }{\beta}\sum_k G(\omega_k)G( (\omega_{1-2})_n - \omega_k)B((\omega_{1-2})_n)
\end{align}
Again, $c_B = 1$ and this yields, 
\be
B((\omega_{1-2})_n) = \frac{\frac{1}{\beta}\sum_k G(\omega_k)G((\omega_{1-2})_n-\omega_k)}{1 + \frac{c_B \lambda_A D }{\beta}\sum_k G(\omega_k)G( (\omega_{1-2})_n-\omega_k) }
\ee
This summation can be computed as follows, 
\be
\frac{1}{\beta} \sum_k G(\omega_k)G( \omega_n- \omega_k ) = \frac{1}{m_1} \frac{{\rm coth} \frac{m_1\beta}{2}}{\omega_n^2 + 4m_1^2} + \frac{\beta \csch^2 \frac{m_1\beta}{2}}{8m_1^2 } \, \delta_{n,0}
\ee
Therefore, separated contributions due to zero mode appear in a bubble chain diagram as well. We can simply write them as follows
\be
B(\omega_n) = \frac{1}{m_1} \frac{ \coth \frac{m_1 \beta}{2} }{\omega_n^2 + m_\sigma^2} + \frac{A}{\tilde{\lambda}_A} \delta_{n,0}\,.
\ee
where $m_\sigma$ is $\beta$-dependent mass,
  \begin{align}
m_\sigma^2 = 4 m_1^2 + \frac{ \tilde{\lambda}_A}{m_1} \coth \frac{m_1 \beta}{2} \,,
  \end{align}
  and $A$ is a complicated $\beta$-dependent function.
  \begin{align}
  \frac{A}{\tilde{\lambda}_A} \,
  = \frac{4 \beta m_1^4 \, }{\left(4 m_1^3 +  \tilde{\lambda}_A \coth \left(\frac{ \beta m_1}{2}\right)\right) \left(4 m_1^3 \left( \cosh \left(\beta m_1 \right) - 1 \right) +  \tilde{\lambda}_A \bigl(\beta m_1 +\sinh (\beta m_1) \bigr)\right)}
  \end{align}
Note that $A$ is positive since $ \tilde{\lambda}_A > 0$.  
  
Since $\lim_{\beta \to \infty} A(\beta) \to 0$, 
\begin{align}
m_\sigma^2 \to 4m_1^2 + \frac{ \tilde{\lambda}_A}{m_1}  \,, \quad
B(\omega_n) \to \frac{1}{m_1} \frac{ 1 }{\omega_n^2 + 4 m_1^2 + \frac{\tilde{\lambda}_A}{m_1}}\,, \quad 
\mbox{as $m_1 \beta \to \infty$} 
\end{align}
Note also that $m_{\sigma}>2$ at any temperature as long as the theory is not free, {\it i.e.,} $\tilde{\lambda}_A>0$. 

We also have  
  \begin{align}
1 - \tilde{\lambda}_A B(\omega_n) = 1 - \frac{\tilde{\lambda}_A}{m_1}\frac{ \coth \frac{m_1 \beta}{2}}{\omega_n^2 + m_\sigma^2} - A \delta_{n,0} = 1 - \frac{m_\sigma^2 - 4m_1^2 }{\omega_n^2 + m_\sigma^2} - A \delta_{n,0} = D_\sigma(\omega_n)
  \end{align}
which matches with $D_\sigma(\omega_n)$ in \eqref{thermalsigma2} up to $\order{1/D}$ corrections coming from the mass difference $m-m_1 = \order{1/D}$.

\subsubsection{$1/D$ corrections for dissipation}
Similarly to the zero temperature case, the Schwinger-Dyson equation can be rewritten as
\begin{align}
\widetilde{G}(\omega_n)^{-1}&= G(\omega_n)^{-1} + \frac{1}{D} \left( \Pi_A (\omega_n) + \Pi_C(\omega_n) \right) + \delta m_A + \delta m_C + O\left(\frac{1}{D^2}\right) \\
 \frac{1}{D}  \Pi_A (\omega_n)  &=
  - \frac{c_A}{D} \frac{\tilde{\lambda}_A^2}{\beta} \sum_k G(\omega_k) B(\omega_k - \omega_n) \\
  \frac{1}{D}  \Pi_C (\omega_n)  &= - \frac{c_C}{D}\frac{\tilde{\lambda}_C^2}{\beta^2} \sum_{k,k'} G(\omega_k)G(\omega_{k'})G(\omega_n-\omega_{k+k'})  
\end{align}
where $c_A = 2$, $c_C= 4$. $\Pi_A$ and $\Pi_C$ contribution are given by Fig.~\ref{fig:1/Dcorrection} and Fig.~\ref{Melon}, respectively. 

Contribution for $\Pi_A$ yields 
  \begin{align}
 \frac{1}{D}  \Pi_A (\omega_n) &=-\frac{c_A}{D} \frac{\tilde{\lambda}_A^2}{\beta}  \sum_k G(\omega_k)  B(\omega_n - \omega_k) \nonumber \\
&= -\frac{c_A}{D} \tilde{\lambda}_A^2 \coth \frac{m_1\beta}{2} 
\frac{m_\sigma \coth \left(\frac{B m_1}{2}\right) \left(-m_1^2+m_\sigma^2+\omega_n^2\right)+m_1   \coth \left(\frac{B m_\sigma}{2}\right) \left(m_1^2-m_\sigma^2+\omega_n^2\right)}{2 m_1^2
   m_\sigma \left((m_1-m_\sigma)^2+\omega_n^2\right) \left((m_1+m_\sigma)^2+\omega_n^2\right)} \nonumber \\
& \quad - \frac{B_1}{\omega_n^2 + m_1^2} \,, 
  \end{align}
  where $c_A=2$ and 
  \be
    \label{B1factor}
  B_1 = \frac{2}{D} \frac{\tilde{\lambda}_A}{\beta} A
  \ee

Contribution for $\Pi_C$ yields, 
  \begin{align}
\frac{1}{D}\Pi_C(\omega_n) &= - \frac{c_C}{D}\frac{\tilde{\lambda}_C^2}{\beta^2} \sum_{k,k'} G(\omega_k)G(\omega_{k'})G(\omega_n-\omega_{k+k'})  \nonumber \\
&= -  \frac{c_C}{D} \frac{\tilde{\lambda}_C^2}{\beta} \sum_k G(\omega_k) \left( \frac{1}{\beta} \sum_{k'} G(\omega_{k'}) G(\omega_n-\omega_{k+k'}) \right) \nonumber \\
&= - \frac{c_C}{D} \frac{\tilde{\lambda}_C^2}{\beta} \sum_k G(\omega_k) \left( \frac{1}{m_1} \frac{{\rm coth} \frac{m_1 \beta}{2}}{\omega_{n-k}^2 + 4m_1^2} + \frac{\beta  \csch^2 \frac{m_1 \beta}{2}}{8m_1^2}\delta_{n-k,0} \right) \nonumber \\
&= - \frac{c_C}{D}\frac{\tilde{\lambda}_C^2}{16m_1^2} \left(  \frac{{\rm csch} ^2 \frac{m_1 \beta}{2}}{\omega_n^2 + m_1^2} + \frac{3(4+3{\rm csch} ^2 \frac{m_1\beta}{2})}{\omega_n^2 + 9m_1^2}\right) - \frac{c_C}{D} \frac{\tilde{\lambda}_C^2}{8m_1^2 } {\rm csch} ^2 \frac{m_1 \beta}{2}  G(\omega_n) \quad \nonumber \\
&= - \frac{B_2}{\omega_n^2 + m_1^2}  - \frac{3 c_C \tilde{\lambda}_C^2}{16D m_1^2}   \frac{4+3{\rm csch} ^2 \frac{m_1 \beta}{2}}{\omega_n^2 + 9m_1^2}   
  \end{align}
where  $c_C = 4$ and 
\begin{align}
\label{B2factor}
B_2 = \frac{3 c_C \tilde{\lambda}_C^2}{16 D m_1^2} \csch^2 \frac{m_1 \beta}{2} =  \frac{3 c_C\tilde{\lambda}_C^2}{4 D m_1^2 }\left( \frac{1}{e^{\beta m_1}-1} \right)^2 e^{ \beta m_1} 
\end{align}
Combining \eqref{B2factor} with \eqref{B1factor}, we obtain the propagator in the vicinity of $\omega^2 + m_1^2 = 0$ as 
\be
\widetilde{G}(\omega_n)^{-1}= G(\omega_n)^{-1} + \frac{1}{D} \left( \Pi_A (\omega_n) + \Pi_C(\omega_n) \right) + \delta m_A + \delta m_C + O\left(\frac{1}{D^2}\right) 
\ee
where 
\begin{align}
& \frac{1}{D} \left( \Pi_A (\omega_n) + \Pi_C(\omega_n) \right) =  - \frac{B}{\omega_n^2 + m_1^2} + \cdots \,, \\
& B = B_1 + B_2 = \frac{1}{D} \left( \frac{2 \tilde{\lambda}_A A}{\beta}  +   \frac{3 \tilde{\lambda}_C^2}{m_1^2 }\left( \frac{1}{e^{\beta m_1}-1} \right)^2 e^{ \beta m_1}  \right) 
\end{align}
This matches the previous results \eqref{thermalX2} and \eqref{Bdef}, neglecting the additional $\order{1/D}$ corrections coming from the mass difference $m-m_1 = \order{1/D}$.
Thus, the rest of the argument that at finite temperature, the single pole at $\omega^2 + m_1^2 = 0$ splits into a pair of nearby poles is the same as section \ref{Hubbard-Stratonovich}.

\section{Conclusions\label{sect:conclusions}}
We studied a simple model of many-matrix quantum mechanics.  From the matrix point of view we worked in the planar limit, sending $N \rightarrow \infty$ first.  Physical quantities can
then be calculated as an expansion in $1/D$, where $D$ is the number of matrices.  This can be thought of as a particular scaling limit of a tensor model.  It can also be thought of as
a non-commutative generalization of an $O(D)$ vector model.

We focused on dissipation at finite temperature, which we extracted from a 2-point correlator.  A curious fact is that dissipation arises at ${\cal O}(1/\sqrt{D})$.
We expect that this is generically true in many-matrix models.  Another curious fact is that at high temperature the leading dissipative effects are due to the
coupling $g_C$, while $g_A$ only makes subleading contributions.  If we set $g_C = 0$ and then
expand the width (\ref{Gammadef}) for high temperatures, then unlike the linear growth (\ref{hiT}) we find that the leading behavior is temperature-independent.
\be
\Gamma\,\big\vert_{\tilde{\lambda}_C = 0} = {\sqrt{2} m \over \pi \sqrt{D}} + {\cal O}(\beta)
\ee

It would be interesting to understand why the couplings $g_C$ and $g_A$ lead to such different behaviors.  As a possible explanation, note that for any $N$ and $D$ the potential
of the model (\ref{SE}) can be written as
\be
V = {1 \over 2} m_0^2 \, {\rm Tr} \, \big(X^i X^i\big) - {1 \over 4} g_C^2 \, {\rm Tr} \, \big([X^i,X^j]^2\big) + {1 \over 2} \big(g_A^2 - g_C^2\big)  \, {\rm Tr} \, \big(X^i X^i X^j X^j\big)
\ee
The mass term dominates near the origin.  The $({\rm commutator})^2$ term is stable, but has flat directions corresponding to commuting matrices.  The last term
is stable if $g_A > g_C$, but if $g_A < g_C$ it makes the potential unstable at large fields.  Thus we see that
\begin{itemize}
\item
For $g_A > g_C$ the model is stable.
\item
For $g_A = g_C$ the quartic potential has flat directions.  The model is stable for $m_0^2 \geq 0$.\footnote{At large $N$ and large $D$, (\ref{mass2}) shows that the model is stable even for $m_0^2 < 0$.
It would be interesting to explore this further and determine the conditions for stability when $m_0^2 < 0$.}
\item
For $g_A < g_C$ the model has an instability at large fields.
\end{itemize}
This behavior appears to be independent of $N$ and $D$.  The instability for $g_A < g_C$ may be related to the different behaviors of the two couplings.\footnote{We thank
V.P.~Nair for suggesting this possibility.}

It would also be interesting to explore higher orders in perturbation theory.
\begin{itemize}
\item
To ${\cal O}(1/D)$, in the retarded propagator (\ref{DR}), we've seen that the pole at $\omega^2 = m^2$
that is present at tree level splits into a pair of poles at $\omega^2 = m_\pm^2$.  Presumably at higher orders in $1/D$ the splitting continues and additional poles develop.  Does the
width $\Gamma$ remain ${\cal O}(1/\sqrt{D})$?
\item
At ${\cal O}(1/D)$ the correlator (\ref{DR2}) undergoes a ``recurrence'' when $t = 4 \pi m / \sqrt{B}$.  As additional poles develop, does the recurrence timescale get longer?
\item
Perhaps one can study the large-order behavior of perturbation theory.  Does the model develop a continuous spectrum?  Is there a
sign of the instability which is present when $g_C > g_A$?  Is the $1/D$ expansion convergent, or can it be re-summed?
\end{itemize}

Another interesting direction is to add a massive vector as a probe \cite{Iizuka:2001cw, Iizuka:2008hg}. It is known that for a single free matrix coupled to a vector, the out-of-time-ordered correlators (OTOCs) for the vector do not grow exponentially in time \cite{Michel:2016kwn, Iizuka:2023owx}. This is because the matrix is free. For the interacting many-matrix model we studied in this paper, the matrices themselves have nontrivial dissipation. Thus, OTOCs for a vector coupled to our interacting matrices might show nontrivial behavior. It would be interesting to investigate this direction further. 

\bigskip
\goodbreak
\centerline{\bf Acknowledgements}
\noindent
We are grateful to Matt Lippert, V.P.\ Nair, Rob Pisarski and Daniel Robbins for valuable discussions.
The work of TA and NI were supported in part by JSPS KAKENHI Grant Number 24K22886(TA), 18K03619(NI). 
The work of NI was also supported by MEXT KAKENHI Grant-in-Aid for Transformative Research Areas A ``Extreme Universe'' No.\ 21H05184.
DK is supported by U.S.\ National Science Foundation grant PHY-2112548.

\appendix
\section{Auxiliary fields\label{appendix:auxiliary}}
We briefly review the introduction of the auxiliary field $\Sigma$ via a Hubbard -- Stratonovich transformation \cite{Zee:2003mt}.
Consider the 1-D integral
\be
Z = \int_{-\infty}^\infty d\phi \, e^{-S_E} \qquad S_E = {1 \over 2} m^2 \phi^2 + {1 \over 2} g^2 \phi^4
\ee
Introduce $1$ in the path integral, represented as
\be
1 = {1 \over \sqrt{2 \pi}} \int_{-\infty}^\infty d\Sigma \, e^{- {1 \over 2} \big(\Sigma - i g \phi^2\big)^2}
\ee
Up to a normalization this leads to
\be
Z = \int d\phi \, d\Sigma \, e^{-S_E} \qquad
S_E = {1 \over 2} m^2 \phi^2 + {1 \over 2} \Sigma^2 - i g \Sigma \phi^2
\ee
This action is analogous to (\ref{SE3}).  Note that in this form both $\phi$ and $\Sigma$ are integrated over real values.  In the literature one sometimes redefines $\Sigma$ to absorb a factor of $i$.

\section{Combinatorics for Hubbard-Stratonovich approach\label{appendix:combinatorics}}
In this appendix we fix the numerical factors that appear in the diagrams of Fig.~\ref{fig:SDX}, including the flipped forms mentioned in the caption.  These factors are displayed in (\ref{DXinv}).  The strategy is simple: we invert
(\ref{DXinv}) and expand in powers of the couplings.  Numerical factors are fixed by matching to ordinary perturbation theory.

In perturbation theory the two-point function is (with no sum on $i$, and with integrals over Euclidean time suppressed)
\be
\langle X^i X^i e^{-S_{\rm int}} \rangle
\ee
The first diagram in Fig.~\ref{fig:SDX}, including its flipped form, comes from
\be
{1 \over 2} g_C^2 \langle X^i \, {\rm Tr} (X^j X^k X^j X^k) \, X^i \rangle
\ee
Restricting to planar diagrams there are 4 possible contractions for the $X^i$ on the left, and 2 possible contractions for the $X^i$ on the right.  There is one closed matrix index loop,
so the diagram comes with a numerical factor
\be
\label{diag1}
{1 \over 2} g_C^2 \cdot 4 \cdot 2 \cdot N = 4 g_C^2 N
\ee
The second diagram in Fig.~\ref{fig:SDX}, including its flipped form, arises at second order in perturbation theory.  It comes from
\be
{1 \over 2!} (i g_A)^2 \langle X^i \, {\rm Tr} (\sigma X^j X^j) \, {\rm Tr} (\sigma X^k X^k) \, X^i \rangle
\ee
Again restricting to planar diagrams, there are two ways to contract the $X$ fields in which the $\sigma$ contraction runs below an $X$ contraction.  There are also two ways to
contract the $X$ fields in which the $\sigma$ contraction runs above an $X$ contraction.  This gives a total of 4 possible contractions.  There is one closed matrix index loop, so the diagram comes with a numerical factor
\be
\label{diag2}
{1 \over 2!} (i g_A)^2 \cdot 4 \cdot N = - 2 g_A^2 N
\ee
The third diagram in Fig.~\ref{fig:SDX} arises at second order in perturbation theory.  It comes from
\be
{1 \over 2!} \Big({1 \over 2} g_C^2\Big)^2 \langle X^i \, {\rm Tr} (X^j X^k X^j X^k) \, {\rm Tr} (X^m X^n X^m X^n) \, X^i \rangle
\ee
For a planar diagram there are 8 possible contractions for the $X^i$ on the left, followed by 4 possible contractions for the $X^i$ on the right.  There are two closed matrix index loops and one closed
vector index loop, so the numerical factor is
\be
\label{diag3}
{1 \over 2!} \Big({1 \over 2} g_C^2\Big)^2 \cdot 8 \cdot 4 \cdot N^2 \cdot D = 4 g_C^4 N^2 D
\ee
The numerical factors (\ref{diag1}), (\ref{diag2}),(\ref{diag3}) appear inside the square brackets in (\ref{DXinv}).

\section{Combinatorics for direct approach\label{appendix:Coefficients}}
In this appendix, we will calculate various combinatoric coefficients in various diagrams\footnote{In this appendix, we use $a, b, c = 1 , \cdots D$ for flavor indices and $i,j, k = 1 , \cdots N$ for color indices.}. 
\subsection{Coefficient $c_L = 2$\label{app:leadingc}}
\begin{figure}[htbp]
    \centering
    \includegraphics[keepaspectratio, scale=0.15]{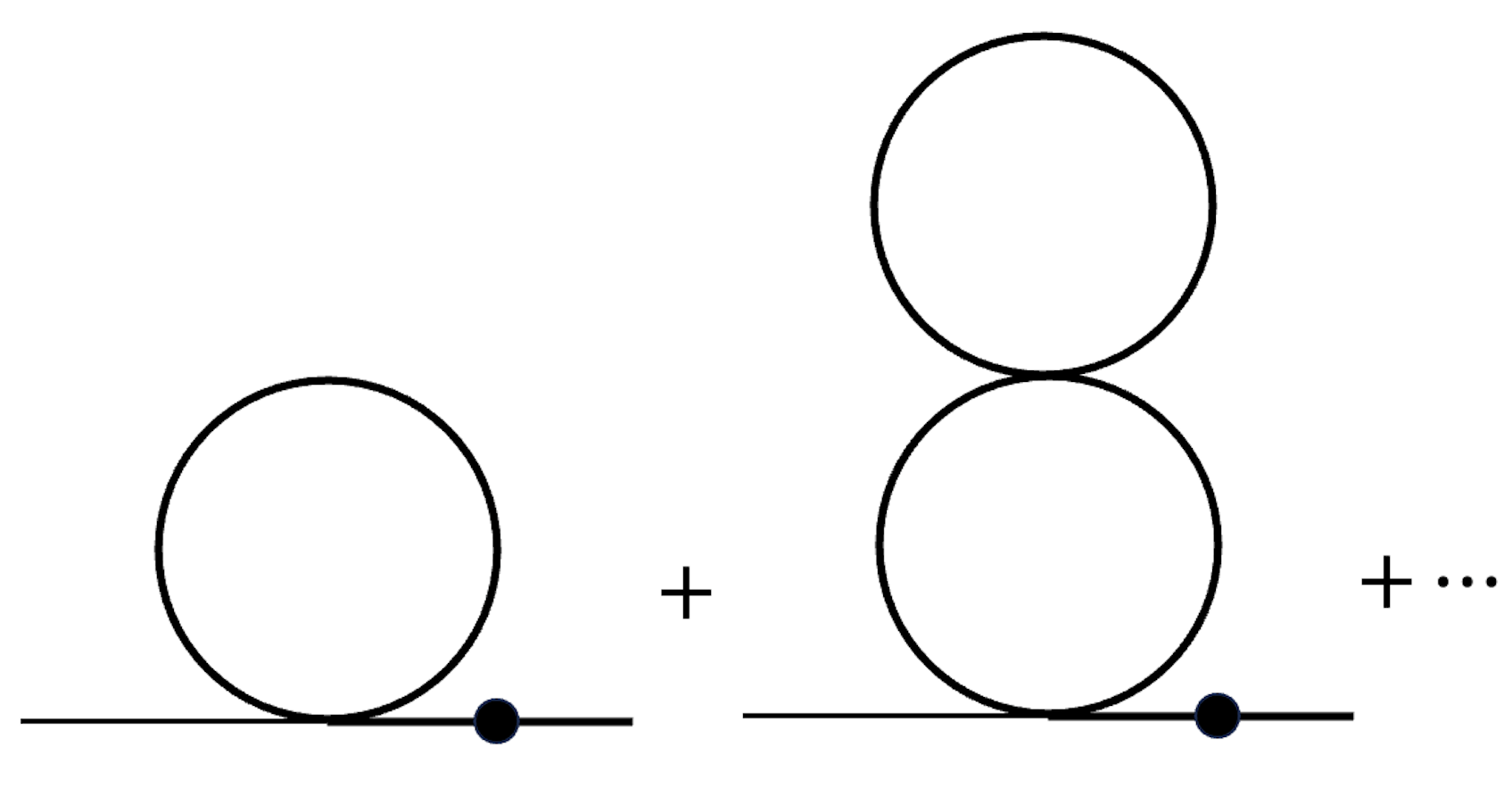}
    \caption{Perturbative snail diagrams to two loops.}
    \label{fig:doublesnail}
\end{figure}

This coefficient can be obtained by Wick contraction as follows. In perturbation theory, the snail diagrams are shown in Fig.~\ref{fig:doublesnail} through two loops.
The one-loop diagram in Fig.~\ref{fig:doublesnail} is 
\bea
X^a_{ij}  \left(- \frac{g_A^2}{2}  \sum_{b,c} \mbox{Tr} \left( X^b X^b X^c X^c \right)\right) X^a_{kl}
\eea
where $X^a$ represents an external line. 
Considering the Trace cyclic property, the possible Wick contractions are as follows.
\begin{align}
&- \frac{ g^2_A}{2} \times 4 \times \sum_{b,c=1}^D  \wick{1}{<x X^a_{ij} >x X^{b}_{mn} X^{b}_{no} X^c_{op} X^c_{pm} X^a_{kl}}  \to -\frac{ g^2_A}{2} \times 4 \times \sum_{b,c=1}^D \wick{121}{ <x X^a_{ij} >x X^b_{mn} <y X^b_{no} <z X^c_{op} >z X^c_{pm} >y X^a_{kl}}  \nonumber \\
&= - 2 g^2_A \sum_{b,c=1}^D \delta^{ab} \delta^{cc} \delta_{in} \delta_{jm} \delta_{nl} \delta_{ok} \delta_{om} \delta_{pp}   = - 2  \tilde{\lambda}_A  \delta_{il} \delta_{jk} 
\end{align}
Here ``$\to $'' means we maintain only the leading contributions and neglect all subleading contributions. 
$4$ is because there are 4 $X$'s in $X^b X^b X^c X^c$. 
This determines 
\be
\label{cL2}
c_L = 2  \,.
\ee
To simplify the notation, let us write only the subscripts of the matrix, such as $a \equiv X^a$. Then 
the two-loop diagram in Fig.~\ref{fig:doublesnail} is 
\begin{align}
& \frac{1}{2!}\left(- \frac{ g_A^2}{2}\right)^2 \sum_{b,c,d,e} X^a (X^b X^b X^c X^c) (X^d X^d X^e X^e) X^a =  \frac{1}{2!}\left(- \frac{ g_A^2}{2}\right)^2 \sum_{b,c,d,e} a (bbcc) (ddee) a \\
& =\frac{(-g_A^2)^2}{2^3} \times 8 \times \sum_{c,d,e} \wick{1}{<x a (>x a a c c) (d d e e) a} 
\to \frac{(-g_A^2)^2}{2^3} \times 8 \times \sum_{c,d,e} \wick{11}{<x a (>x a <y a c c) (d d e e) >y a}\\ 
&\to \frac{(-g_A^2)^2}{2^3} \times 8 \times 2 \times \sum_{c,d} \wick{14321}{<x a (>x a <y a <zc <a c) (<b d >b d >a c >zc) >y a} \, + \,  \frac{(-g_A^2)^2}{2^3} \times 8 \times 2 \times \sum_{c,d} \wick{14321}{<x a (>x a <y a <a c <zc) (<b d >b d >a c >zc) >y a}
\end{align}
$8$ is because there are 8 $X$'s in $bbccddee$ and $2$ is because two choices $d$ and $e$ to pair up with $c$.  
The last term is also planar due to the trace cyclicity. Thus in this two-loop diagram, the final coefficient is $(-2 \lambda_A D)^2$. This also justifies \eqref{cL2}.

\subsection{Coefficient $c_B = 1$ in the bubble chain diagram\label{app:bubblec}}
The coefficient $c_B$ in a bubble diagram can be obtained similarly. 
First, we will calculate the coefficients in the perturbative bubble diagram as 
\begin{figure}[htbp]
    \centering
    \includegraphics[keepaspectratio, scale=0.6]{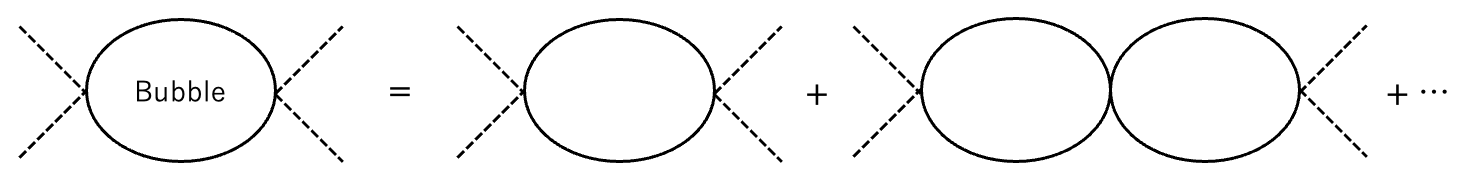}
    \caption{Bubble chain diagrams}
    \label{app:bubble}
\end{figure}

The first term on the right-hand side of Fig.~\ref{app:bubble} is 
  \bea
 \frac{1}{2!} \left( -\frac{ g_A^2}{2}\right)^2 \sum_{a,b,c,d} X^I X^I(X^a X^a X^b X^b)(X^c X^c X^d X^d) X^F X^F
  \eea
and in the simplified notation, only the following wick contraction needs to be considered if we distinguish between two external lines $I$'s and the others $F$'s.
  \bea
\frac{(-g_A^2)^2}{2^3} \times 8 \times 4 \times \wick{2121}{<1 I <2 I(>2 a >1 a b b)( cc <3 d <4 d)>4 F >3 F}
  \eea
$8$ is because there are 8 $X$'s in $aabbccdd$. $4$ is to choose one in $ccdd$.    

Then there are two choices to contract between $b$ and $c$'s. 
  \begin{align}
\frac{(-g_A^2)^2}{2^3} \times 8 \times 4 \times  \wick{212121}{<1 I <2 I(>2 a >1 a <5 b <6 b)( >6 c >5 c <3 d <4 d)>4 F >3 F}\\
\frac{(-g_A^2)^2}{2^3} \times 8 \times 4 \times  \wick{212121}{<1 I <2 I(>2 a >1 a <5 b <6 b)( >5 c >6 c <3 d <4 d)>4 F >3 F}
  \end{align}
The first one is a planar diagram,  
  \begin{align}
&\wick{212121}{<1 I_{ij} <2 I_{i'j'}(>2 a_{mn} >1 a_{no} <5 b_{op} <6 b_{pm})( >6 c_{qr} >5 c_{rs} <3 d_{st} <4 d_{tq})>4 F_{kl} >3 F_{k'l'}} \nonumber \\
& = \sum_{m,n,o,p,q,r,s,t} \delta_{io}\delta_{jn} \cdot \delta_{i'n}\delta_{j'm} \cdot \delta_{os}\delta_{pr} \cdot \delta_{pr}\delta_{mq} \cdot  \delta_{sl'}\delta_{tk'} \cdot \delta_{tl}\delta_{qk} = N \delta_{il'}\delta_{ji'} \delta_{j'k}\delta_{lk'}
  \end{align}
  but the second one is non-planar suppressed by $1/N$, 
  \begin{align}
&\wick{212121}{<1 I_{ij} <2 I_{i'j'}(>2 a_{mn} >1 a_{no} <5 b_{op} <6 b_{pm})( >5 c_{qr} >6 c_{rs} <3 d_{st} <4 d_{tq})>4 F_{kl} >3 F_{k'l'}}\\
& = \sum_{m,n,o,p,q,r,s,t} \delta_{io}\delta_{jn} \cdot \delta_{i'n}\delta_{j'm} \cdot \delta_{or}\delta_{pq} \cdot \delta_{ps}\delta_{mr} \cdot \delta_{sl'}\delta_{tk'} \cdot \delta_{tl}\delta_{qk} = \delta_{ij'}\delta_{ji'}\delta_{lk'} \delta_{kl'}
  \end{align}
Thus, the leading coefficient is
  \bea
  \label{lambdaA2bubble}
\frac{(-g_A^2)^2 N}{2^3} \times 8 \times 4 = 4 \times  \frac{(- \lambda_A D)^2}{ND}
  \eea

The second term on the right-hand side of Fig.~\ref{app:bubble} is 
  \bea
 \frac{1}{3!} \left( - \frac{g_A^2}{2}\right)^3 \sum_{a,b,c,d,e,g} II(aabb)(ccdd)(eegg)FF
  \eea
The contractions of $I$ and $F$, respectively, can be written as follows.
  \bea
12 \cdot 8 \cdot \wick{2121}{<1I<2I(>2a>1abb)(ccdd)(ee<3g<4g)>4F>3F}
  \eea
  12 is because 12 $X$'s in $aabbccddeegg$ and 8 is because 8 $X$'s in $ccddeegg$. 
  
And there are four ways to contract, 
  \bea
\label{app:bub3}
&&12 \cdot 8 \cdot 2 \cdot \wick{21212121}{<1I<2I(>2a>1a <5 b <6 b)(>6 c >5 c <7 d <8 d)(>8 e >7 e<3g<4g)>4F>3F}\\
\label{app:bub4}
&&12 \cdot 8 \cdot 2 \cdot \wick{21212121}{<1I<2I(>2a>1a <5 b <6 b)(>6 c >5 c <7 d <8 d)(>7 e >8 e<3g<4g)>4F>3F}\\
\label{app:bub5}
&&12 \cdot 8 \cdot 2 \cdot \wick{21212121}{<1I<2I(>2a>1a <5 b <6 b)(>5 c >6 c <7 d <8 d)(>8 e >7 e<3g<4g)>4F>3F}\\
\label{app:bub6}
&&12 \cdot 8 \cdot 2 \cdot \wick{21212121}{<1I<2I(>2a>1a <5 b <6 b)(>5 c >6 c <7 d <8 d)(>7 e >8 e<3g<4g)>4F>3F}
  \eea
  2 is because of two choices between $c$ and $d$. 
However, one can check that only \eqref{app:bub3} is a planar,
\begin{align}
&\wick{21212121}{<1I_{ij}<2I_{i'j'}(>2a_{mn}>1a_{no} <5 b_{op} <6 b_{pm})(>6 c_{qr} >5 c_{rs} <7 d_{st} <8 d_{tq})(>8 e_{uv} >7 e_{vw}<3g_{wx}<4g_{xu})>4F_{kl}>3F_{k'l'}}\\
& = \sum_{m,n,o,p,q,r,s,t,u,v,w,x} \delta_{io}\delta_{jn} \cdot \delta_{i'n}\delta_{j'm}  \cdot \delta_{os}\delta_{pr}  \cdot  \delta_{pr}\delta_{mq}  \cdot \delta_{sw}\delta_{tv}  \cdot \delta_{tv}\delta_{qu}  \cdot  \delta_{wl'}\delta_{xk'}  \cdot \delta_{xl}\delta_{uk} \\
&=  N^2 \delta_{il'}\delta_{ji'} \delta_{j'k} \delta_{lk'}
\end{align}
and the rest are non-planar. 
Thus, the leading coefficient is
  \bea
  \label{lambdaA3bubble}
 \frac{1}{3!} \left(- \frac{g_A^2}{2}\right)^3 \times 12 \times 8 \times 2= 4 \times \frac{(-\lambda_A D)^3}{ND}
  \eea
  
Comparison between \eqref{lambdaA2bubble} and \eqref{lambdaA3bubble} gives   
\be
c_B=1 \,.
\ee 
for the bubble chain contribution given by the geometry sum as \eqref{geometricformBomega}.

\subsection{Coefficient $c_A = 2$ for Fig.~\ref{fig:1/Dcorrection} \label{app:typea}}
\begin{figure}[htbp]
 \qquad  \qquad  \quad \begin{minipage}[b]{0.4\linewidth}
    \includegraphics[keepaspectratio, scale=0.51]{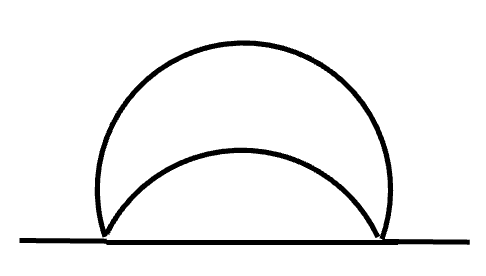}
    \caption{One-loop bubble diagram, proportional to $2(-\lambda_A D)^2 \frac{1}{D}$.}
    \label{fig:typea-1}
  \end{minipage}
  \qquad
  \begin{minipage}[b]{0.4\linewidth}
    \centering
    \includegraphics[keepaspectratio, scale=0.39]{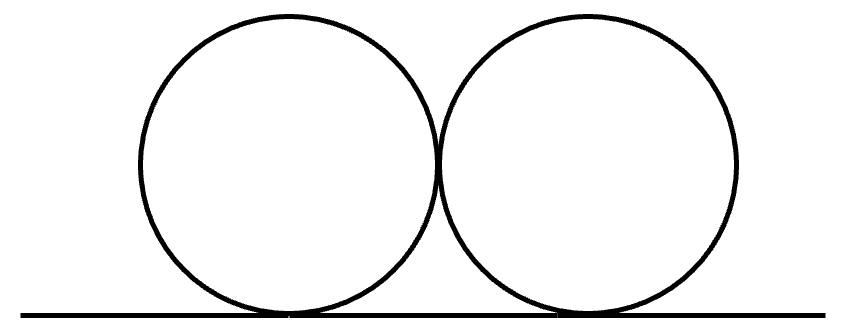}
    \caption{Two-loop bubble diagram, proportional to $2(-\lambda_A D)^3 \frac{1}{D}$.}
    \label{fig:typea-2}
  \end{minipage}
\end{figure}

We would like to consider the contribution from resumed bubbles as the Fig.~\ref{fig:1/Dcorrection}. For that purpose, let us first consider those with just one loop or two loops as Fig.~\ref{fig:typea-1} and \ref{fig:typea-2}.

\noindent Let us consider Fig. \ref{fig:typea-1} first. This is obtained from the term 
\begin{align}
\frac{1}{2!}\left(- \frac{ g_A^2}{2}\right)^2 \sum_{b,c,d,e} a (b b c c) (d d e e) a
\end{align}
by the following Wick contractions, 
\begin{align}
\label{typea:1-1}
4 \times 2 \times \frac{(-g_A^2)^2}{2^3}   \sum_{b,c,d,e} \wick{14321}{ <1 a ( >1 b <2 b <3 c <4 c) <5 a (>5 d  >4 e >3 e >2 d) }
\quad \mbox{or} \quad 4 \times 2 \times  \frac{(-g_A^2)^2}{2^3}  \sum_{b,c,d,e} \wick{13211}{ <1 a ( >1 b <2 c <3 c <4 b) (>4 d  >3 e >2 e <5 d) >5 a }
\end{align}
4 is for four choices in $bcde$ and 2 is for two choices in $de$, and the rest of contractions are subleading.

From these two, we obtain 
\be
4 \times 2 \times  \frac{(-g_A^2)^2}{2^3}  \times 2 \times N^2 D =  2  \frac{(-\lambda_A D)^2}{D}
\ee

Next, let us consider Fig. \ref{fig:typea-2}. 
\be
\frac{1}{3!}\left(-\frac{ g_A^2}{2}\right)^3 \sum_{b,c,d,e,f,g} a (b b c c) (d d e e) (f f g g) a  
\ee
We have the following Wick contractions in the leading order,  
\begin{align}
&6 \times 4 \times 2 \times \frac{1}{3!}\left(-\frac{ g_A^2}{2}\right)^3 \sum_{b,c,d,e,f,g}  \wick{1321211}{ <1 a ( >1 b <2 b <4 c <5 c) (>5 d >4 d <6 e <7 e) (>7 f >6 f >2 g <3 g) >3 a }
\label{typea:2-2}\\
\mbox{or} \quad 
& 6 \times 4 \times 2 \times \frac{1}{3!}\left(-\frac{ g_A^2}{2}\right)^3 \sum_{b,c,d,e,f,g}  \wick{1541321}{ <1 a ( >1 b <4 c <5 c <2 b)(>2 g <6 f <7 f <3 g) >3 a  (>7 e >6 e >5 d >4 d) }  
\end{align}
6 is for six choices in $bcdefg$ and 4 is for four choices in $defg$. 2 is for two choices in $de$. 
Therefore, the coefficient is determined as 
\be
 6 \times 4 \times 2 \times \frac{1}{3!}\left(- \frac{ g_A^2}{2}\right)^3 \times 2 \times N^3 D^2 = 2 \frac{(-\lambda_A D)^3}{D}
\ee

From the above, we see that the overall coefficients of both Fig.~\ref{fig:typea-1} and \ref{fig:typea-2} is 2. 
Thus we can write down these diagram contributions as
\begin{align}
& \quad \frac{2}{D} (- \lambda_A D)^2 G_0(\omega)G(\omega) \int \frac{d\omega_1}{2\pi} G(\omega_1)
 \left[ \int \frac{d\omega'}{2\pi} G(\omega')G(\omega'-\omega + \omega_1)\right.  \nonumber \\
&\ \ \ \ \ \ \left. + \, c_B(- \lambda_A D) \int \frac{d\omega'}{2\pi} G(\omega')G(\omega'-\omega + \omega_1) \int \frac{d\omega''}{2\pi} G(\omega'')G(\omega''-\omega + \omega_1) +\cdots \right]  \nonumber \\
&= \frac{2}{D} (- \lambda_A D)^2 G_0(\omega)G(\omega) \int \frac{d\omega_1}{2\pi} G(\omega_1) B(\omega - \omega_1)
\end{align}
with 
\be
c_A = 2 \,.
\ee

\subsection{Coefficient $c_C = 4$\label{app:typec}}
Let us consider the melon diagram, which plays an important role in the two-point function.
\begin{figure}[h]
\centering
\includegraphics[keepaspectratio,scale=0.65]{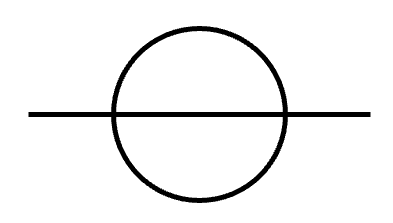}
\caption{Melon diagram proportional to $\lambda_C$.}
\label{fig:typc}
\end{figure}
From Fig.~\ref{fig:typc}, this can be obtained from
\be
 \frac{1}{2!} \left( \frac{g_C^2}{2}\right)^2 \sum_{b,c,d,e} a(bcbc)(dede)a
\ee
From the cyclic property, all way of contraction can be written as
\be
8 \times 4 \times   \frac{1}{2!} \left( \frac{g_C^2}{2}\right)^2 \sum_{b,c,d,e} \wick{11}{<1 a( >1 bcbc)(ded< 2e) >2 a}
\ee
8 is because of 8 $X$'s in $bcbcdede$ and 4 is because of 4 $X$'s in $dede$. 
From this, there is only one way to contract in the leading order as 
  \be
8 \times 4 \times \wick{13211}{<1 a( >1 b <3 c <4 b <5 c)(>5 d >4 e >3 d< 2e) >2 a}
  \ee
Therefore, the coefficient is
\be
8 \times 4 \times  \frac{1}{2!} \left( \frac{g_C^2}{2}\right)^2 \times N^2 D  = 4 \frac{(\lambda_C D)^2}{D} 
\ee
Thus, 
\be
c_C = 4 \,.
\ee


\begin{thebibliography}{10}

\bibitem{Banks:1996vh}
T.~Banks, W.~Fischler, S.~H. Shenker, and L.~Susskind, ``{M theory as a matrix
  model: A conjecture},''
  \href{http://dx.doi.org/10.1103/PhysRevD.55.5112}{{\em Phys. Rev. D}
  {\bfseries 55} (1997) 5112--5128},
  \href{http://arxiv.org/abs/hep-th/9610043}{{\ttfamily arXiv:hep-th/9610043}}.

\bibitem{Klebanov:2018fzb}
I.~R. Klebanov, F.~Popov, and G.~Tarnopolsky, ``{TASI lectures on large $N$
  tensor models},'' \href{http://dx.doi.org/10.22323/1.305.0004}{{\em PoS}
  {\bfseries TASI2017} (2018) 004},
  \href{http://arxiv.org/abs/1808.09434}{{\ttfamily arXiv:1808.09434
  [hep-th]}}.

\bibitem{Ferrari:2017ryl}
F.~Ferrari, ``{The large $D$ limit of planar diagrams},''
  \href{http://dx.doi.org/10.4171/aihpd/76}{{\em Ann. Inst. H. Poincare D Comb.
  Phys. Interact.} {\bfseries 6} no.~3, (2019) 427--448},
  \href{http://arxiv.org/abs/1701.01171}{{\ttfamily arXiv:1701.01171
  [hep-th]}}.
  
\bibitem{Azeyanagi:2017drg}
T.~Azeyanagi, F.~Ferrari, and F.~I. Schaposnik~Massolo, ``{Phase diagram of
  planar matrix quantum mechanics, tensor, and Sachdev-Ye-Kitaev models},''
  \href{http://dx.doi.org/10.1103/PhysRevLett.120.061602}{{\em Phys. Rev.
  Lett.} {\bfseries 120} no.~6, (2018) 061602},
  \href{http://arxiv.org/abs/1707.03431}{{\ttfamily arXiv:1707.03431
  [hep-th]}}.

\bibitem{Hotta:1998en}
T.~Hotta, J.~Nishimura, and A.~Tsuchiya, ``{Dynamical aspects of large $N$
  reduced models},''
  \href{http://dx.doi.org/10.1016/S0550-3213(99)00056-5}{{\em Nucl. Phys. B}
  {\bfseries 545} (1999) 543--575},
  \href{http://arxiv.org/abs/hep-th/9811220}{{\ttfamily arXiv:hep-th/9811220}}.

\bibitem{Mandal:2009vz}
G.~Mandal, M.~Mahato, and T.~Morita, ``{Phases of one dimensional large $N$ gauge
  theory in a $1/D$ expansion},''
  \href{http://dx.doi.org/10.1007/JHEP02(2010)034}{{\em JHEP} {\bfseries 02}
  (2010) 034}, \href{http://arxiv.org/abs/0910.4526}{{\ttfamily arXiv:0910.4526
  [hep-th]}}.

\bibitem{Kolganov:2022mpe}
N.~Kolganov and D.~A. Trunin, ``{Classical and quantum butterfly effect in
  nonlinear vector mechanics},''
  \href{http://dx.doi.org/10.1103/PhysRevD.106.025003}{{\em Phys. Rev. D}
  {\bfseries 106} no.~2, (2022) 025003},
  \href{http://arxiv.org/abs/2205.05663}{{\ttfamily arXiv:2205.05663
  [hep-th]}}.

\bibitem{Stanford:2015owe}
D.~Stanford, ``{Many-body chaos at weak coupling},''
  \href{http://dx.doi.org/10.1007/JHEP10(2016)009}{{\em JHEP} {\bfseries 10}
  (2016) 009}, \href{http://arxiv.org/abs/1512.07687}{{\ttfamily
  arXiv:1512.07687 [hep-th]}}.

\bibitem{Bjorken:1965sts}
J.~D. Bjorken and S.~D. Drell, {\em {Relativistic Quantum Mechanics}}.
\newblock International Series In Pure and Applied Physics. McGraw-Hill, New
  York, 1965.
\newblock See sections 19.2 and 19.3.

\bibitem{Roberts:1994dr}
C.~D. Roberts and A.~G. Williams, ``{Dyson-Schwinger equations and their
  application to hadronic physics},''
  \href{http://dx.doi.org/10.1016/0146-6410(94)90049-3}{{\em Prog. Part. Nucl.
  Phys.} {\bfseries 33} (1994) 477--575},
  \href{http://arxiv.org/abs/hep-ph/9403224}{{\ttfamily arXiv:hep-ph/9403224}}.

\bibitem{vonSmekal:1997ern}
L.~von Smekal, A.~Hauck, and R.~Alkofer, ``{A solution to coupled
  Dyson\textendash{}Schwinger equations for gluons and ghosts in Landau
  gauge},'' \href{http://dx.doi.org/10.1006/aphy.1998.5806}{{\em Annals Phys.}
  {\bfseries 267} (1998) 1--60},
  \href{http://arxiv.org/abs/hep-ph/9707327}{{\ttfamily arXiv:hep-ph/9707327}}.
  [Erratum: Annals Phys. 269, 182 (1998)].

\bibitem{Kabat:1999hp}
D.~N. Kabat and G.~Lifschytz, ``{Approximations for strongly coupled
  supersymmetric quantum mechanics},''
  \href{http://dx.doi.org/10.1016/S0550-3213(99)00818-4}{{\em Nucl. Phys. B}
  {\bfseries 571} (2000) 419--456},
  \href{http://arxiv.org/abs/hep-th/9910001}{{\ttfamily arXiv:hep-th/9910001}}.

\bibitem{Pisarski:1987wc}
R.~D. Pisarski, ``{Computing finite temperature loops with ease},''
  \href{http://dx.doi.org/10.1016/0550-3213(88)90454-3}{{\em Nucl. Phys. B}
  {\bfseries 309} (1988) 476--492}.

\bibitem{Parwani:1991gq}
R.~R. Parwani, ``{Resummation in a hot scalar field theory},''
  \href{http://dx.doi.org/10.1103/PhysRevD.45.4695}{{\em Phys. Rev. D}
  {\bfseries 45} (1992) 4695},
  \href{http://arxiv.org/abs/hep-ph/9204216}{{\ttfamily arXiv:hep-ph/9204216}}.
  [Erratum: Phys.Rev.D 48, 5965 (1993)].

\bibitem{Laine:2016hma}
M.~Laine and A.~Vuorinen,
  \href{http://dx.doi.org/10.1007/978-3-319-31933-9}{{\em {Basics of thermal
  field theory}}}, vol.~925.
\newblock Springer, 2016.
\newblock \href{http://arxiv.org/abs/1701.01554}{{\ttfamily arXiv:1701.01554
  [hep-ph]}}.

\bibitem{2001CMaPh.216...59C}
G.~{Cuniberti}, E.~{De Micheli}, and G.~A. {Viano}, ``{Reconstructing the
  thermal Green functions at real times from those at imaginary times},''
  \href{http://dx.doi.org/10.1007/s002200000324}{{\em Communications in
  Mathematical Physics} {\bfseries 216} no.~1, (Jan., 2001) 59--83},
  \href{http://arxiv.org/abs/cond-mat/0109175}{{\ttfamily
  arXiv:cond-mat/0109175 [cond-mat.str-el]}}.

\bibitem{Iizuka:2001cw}
N.~Iizuka, D.~N. Kabat, G.~Lifschytz, and D.~A. Lowe, ``{Probing black holes in
  nonperturbative gauge theory},''
  \href{http://dx.doi.org/10.1103/PhysRevD.65.024012}{{\em Phys. Rev. D}
  {\bfseries 65} (2002) 024012},
  \href{http://arxiv.org/abs/hep-th/0108006}{{\ttfamily arXiv:hep-th/0108006}}.

\bibitem{Iizuka:2008hg}
N.~Iizuka and J.~Polchinski, ``{A matrix model for black hole
  thermalization},''
  \href{http://dx.doi.org/10.1088/1126-6708/2008/10/028}{{\em JHEP} {\bfseries
  10} (2008) 028}, \href{http://arxiv.org/abs/0801.3657}{{\ttfamily
  arXiv:0801.3657 [hep-th]}}.

\bibitem{Michel:2016kwn}
B.~Michel, J.~Polchinski, V.~Rosenhaus, and S.~J. Suh, ``{Four-point function
  in the IOP matrix model},''
  \href{http://dx.doi.org/10.1007/JHEP05(2016)048}{{\em JHEP} {\bfseries 05}
  (2016) 048}, \href{http://arxiv.org/abs/1602.06422}{{\ttfamily
  arXiv:1602.06422 [hep-th]}}.

\bibitem{Iizuka:2023owx}
N.~Iizuka and M.~Nishida, ``{Out-of-time-ordered correlators in the IP matrix
  model},'' \href{http://dx.doi.org/10.1007/JHEP05(2024)026}{{\em JHEP}
  {\bfseries 05} (2024) 026}, \href{http://arxiv.org/abs/2311.18137}{{\ttfamily
  arXiv:2311.18137 [hep-th]}}.

\bibitem{Zee:2003mt}
A.~Zee, {\em {Quantum Field Theory in a Nutshell}}.
\newblock Princeton University Press, 2003.
\newblock See page 192.

\end{thebibliography}
\providecommand{\href}[2]{#2}\begingroup\raggedright\endgroup

\end{document}